\documentclass[prd,preprintnumbers,twocolumn,amsmath,nofootinbib,amssymb]{revtex4}
\usepackage{graphicx,color,dcolumn,booktabs,bm}
\usepackage{longtable,lscape}
\usepackage{txfonts}
\usepackage{overpic}
\usepackage{amssymb}
\usepackage{epstopdf}
\usepackage{indentfirst}
\usepackage{feynmf}   
\usepackage{slashed}  
\usepackage{cases}
\usepackage{color}
\usepackage{float}
\usepackage{multirow}
\usepackage{ulem}
\usepackage{graphicx,color,dcolumn,booktabs,bm}
\usepackage{epsfig,dsfont,amssymb,amsmath,amsfonts,amsbsy,mathrsfs}

\graphicspath{{Figures/}} %

\usepackage{hyperref}
\hypersetup{colorlinks,citecolor=blue,anchorcolor=red,menucolor=red, linkcolor=red,filecolor=red,runcolor=red,urlcolor=blue,frenchlinks=red}

\makeatletter
\@addtoreset{equation}{section}
\makeatother

\allowdisplaybreaks

\begin{document}

\title{Mass behavior of hidden-charm pentaquarks with open-strange inspired by these established $P_c$ molecular states}

\author{Rui Chen$^{1}$}\email{chenrui@hunnu.edu.cn}
\author{Xiang Liu$^{2,3,4}$}
\email{xiangliu@lzu.edu.cn}

\affiliation{
$^1$Key Laboratory of Low-Dimensional Quantum Structures and Quantum Control of Ministry of Education, Department of Physics and Synergetic Innovation Center for Quantum Effects and Applications, Hunan Normal University, Changsha 410081, China\\
$^2$Lanzhou Center for Theoretical Physics, Key Laboratory of Theoretical Physics of Gansu Province, and Frontier Science Center for Rare Isotopes, Lanzhou University, Lanzhou 730000, China\\
$^3$School of Physical Science and Technology, Lanzhou University, Lanzhou 730000, China\\
$^4$Research Center for Hadron and CSR Physics, Lanzhou University
and Institute of Modern Physics of CAS, Lanzhou 730000, China}
\date{\today}

\begin{abstract}
  Stimulated by the meson-baryon molecular interpretations of the $P_c$ states ($P_c(4312)/P_c(4440)/P_c(4457)$), we systematically study the interactions between an $S-$wave charm-strange baryon $\Xi_c^{(\prime,*)}$ and an anti-charmed meson $\bar{D}^{(*)}$ in a coupled channel analysis. Effective potentials for the $\Xi_c^{(\prime,*)}\bar{D}^{(*)}$ interactions in a one-boson-exchange model can be related to those in the $\Sigma_c^{(*)}\bar{D}^{(*)}$ systems by using the $SU(3)$ flavor symmetry and heavy quark symmetry. Our results can predict several promising hidden-charm molecular pentaquarks with strangeness $|S|=1$, which include the $\Xi_c^{\prime}\bar{D}$ states with $I(J^P)=0,1(1/2^-)$, the $\Xi_c^*\bar{D}$ states with $0,1(3/2^-)$, the $\Xi_c^{\prime}\bar{D}^*$ states with $0(1/2^-)$ and $0,1(3/2^-)$, and the $\Xi_c^*\bar{D}^*$ states with $0(1/2^-,3/2^-,5/2^-)$.
\end{abstract}


\maketitle

\section{Introduction}\label{sec1}

In 2015, the LHCb Collaboration reported two hidden-charm pentaquarks namely $P_c(4380)$ and $P_c(4450)$ in the $\Lambda_b^0\to J/\psi pK^-$ decay process \cite{Aaij:2015tga}. Their observations immediately inspired theorists propose several different interpretations of the $P_c(4380)$ and $P_c(4450)$ states, i.e., the molecular state assignments \cite{Chen:2015loa,Chen:2015moa,Karliner:2015ina,Roca:2015dva,
Mironov:2015ica,He:2015cea,Meissner:2015mza,Burns:2015dwa,Shimizu:2016rrd,Chen:2016heh,
Eides:2015dtr,Huang:2015uda,Chen:2016otp,Yang:2015bmv,He:2016pfa,Yamaguchi:2016ote}, the diquark-diquark-antiquark configuration \cite{Maiani:2015vwa,Li:2015gta,Ghosh:2015ksa,Anisovich:2015zqa,Wang:2015ava}, the diquark-triquark configuration \cite{Lebed:2015tna,Zhu:2015bba}, the re-scattering effect \cite{Guo:2015umn,Liu:2015fea,Mikhasenko:2015vca}, and so on (see reviews \cite{Chen:2016qju,Liu:2019zoy,Brambilla:2019esw,Guo:2017jvc,Esposito:2016noz,Hosaka:2016pey} for details). Because the $P_c(4380)$ and $P_c(4450)$ are very close to the mass threshold of a charmed baryon and an anti-charmed meson, the hadronic molecular state assignments to the $P_c$ states are the most popular proposal, which already had been predicted previously \cite{Yang:2011wz,Wu:2010jy,Wang:2011rga,Karliner:2015ina,Li:2014gra,Wu:2010vk}.

In 2019, the LHCb Collaboration updated the observations of the $P_c$ states in the same process with data collected in run 1 and run 2 \cite{Aaij:2019vzc}. They discovered three narrow structures ($P_c(4312)^+$, $P_c(4440)^+$, and $P_c(4457)^+$) in the $J/\psi p$ invariant mass spectrum, where the $P_c(4440)^+$ and $P_c(4457)^+$ correspond to the fine structures of the former $P_c(4450)$ \cite{Aaij:2015tga}. The refinement observations of the $P_c$ states may provide a strong evidence of the hidden-charm molecular pentaquarks \cite{Chen:2019asm,Liu:2019tjn,He:2019ify,
Meng:2019ilv,Burns:2019iih,Anwar:2018bpu}, although there were the other discussions on these three $P_c$ states \cite{Weng:2019ynv,Ali:2019npk,Giron:2019bcs,Cheng:2019obk,Mutuk:2019snd}. In particular, the $P_c(4312)$, $P_c(4440)$ and $P_c(4457)$ can be assigned as the loosely bound $\Sigma_c\bar{D}$ state with $(I=1/2,J^P=1/2^-)$, the $\Sigma_c\bar{D}^*$ state with $(I=1/2,J^P=1/2^-)$ and the $\Sigma_c\bar{D}^*$ state with $(I=1/2,J^P=3/2^-)$, respectively, based on the one-boson-exchange potentials with considering the coupled-channel effect \cite{Chen:2019asm}. 

Stimulated by the hidden-charm molecular assignments to the $P_c$ states, many groups further search for the possible hidden-charm molecular pentaquarks with strangeness $|S|=1$ \cite{Chen:2016ryt,Wang:2019nvm,Wang:2015wsa,Feijoo:2015kts,
Lu:2016roh,Xiao:2019gjd,Zhang:2020cdi,Shen:2020gpw,Ferretti:2020ewe,Zhu:2021lhd,Xiao:2021rgp}, $|S|=2$ \cite{Wang:2020bjt}, and $|S|=3$ \cite{Wang:2021hql}. For example, in Ref. \cite{Chen:2016ryt}, we study the single $\Lambda_c\bar{D}_s^*$, $\Sigma_c\bar{D}_s^*$, $\Sigma_c^*\bar{D}_s^*$, $\Xi_c\bar{D}^*$, $\Xi_c'\bar{D}^*$, and $\Xi_c^*\bar{D}^*$ interactions by considering the one-eta-exchange and/or one-pion-exchange processes, and predict several possible strange hidden-charm molecular pentaquarks.

Experimental, very recently, the LHCb Collaboration further analyzed the $\Xi_b^-\to J/\psi\Lambda K^-$ decay process, and reported an evidence of the strange hidden-charm pentaquark $P_{cs}(4459)$ in the $J/\psi\Lambda$ invariant mass spectrum \cite{LHCb:2020jpq}. Its resonance parameters are
\begin{eqnarray*}
M=4458.8\pm2.9_{-1.2}^{+4.7} \text{MeV}, \quad \Gamma= 17.3\pm6.5_{-5.7}^{+8.0} \text{MeV},
\end{eqnarray*}
respectively. Its spin-parity is not determined, yet. Since its mass is just below the $\Xi_c\bar{D}^*$ threshold with around 19 MeV, the $P_{cs}(4459)$ was explained as a strange hidden-charm $\Xi_c\bar{D}^*$ molecule \cite{Chen:2020kco,Chen:2021tip,Chen:2020uif,Peng:2020hql,Wang:2020eep}.

All these $P_c$ and $P_{cs}$ states are a little below the mass thresholds of a pair of a charmed/or charm-strange baryon and an anti-charmed meson. If they are the hidden-charm molecules, in this line, one can expect the existence of more hidden-charm molecular pentaquarks based on the heavy quark symmetry and the $SU(3)$ flavor symmetry. In this work, we will perform a systematic investigation on the $\Xi_c^{(',*)}\bar{D}^{(*)}$ interactions, and we still adopt the one-boson-exchange (OBE) model, including the $\pi$, $\sigma$, $\eta$, $\rho$, and $\omega$ exchanges. The corresponding OBE effective potentials for the $\Xi_c^{(',*)}\bar{D}^{(*)}$ systems can be deduced from the $\Sigma_c^{(*)}\bar{D}^{(*)}$ interactions by using the heavy quark symmetry and the $SU(3)$ flavor symmetry. Here, we also consider the coupled channel effect in our calculations. It is very important in generating the $P_c$ and $P_{cs}$ states as hidden-charm molecular pentaquarks \cite{Chen:2019asm,Chen:2020kco,Chen:2021tip}. By revisiting the $\Xi_c^{(',*)}\bar{D}^{(*)}$ interactions, one can provide valuable information to search for the possible hidden-charm molecular pentaquarks with strangeness $|S|=1$. Our comprehensive investigation can also help us to probe the inner structures or underly mechanism of the $P_c$ and $P_{cs}$ states.

This paper is organized as follows. After introduction, we deduce the OBE effective potentials in Sec. \ref{sec2}. In Sec. \ref{sec3}, we present the corresponding numerical results. The paper ends with a summery in Sec. \ref{sec4}.

\section{OBE effective potentials}\label{sec2}


According to the chiral symmetry, the effective Lagrangians of the interactions between light quarks $(u,d,s)$ and the light mesons ($\sigma$, $\pi$, $\eta$, $\rho$ and $\omega$) are constructed as
\begin{eqnarray}
\mathcal{L}_q &=& -g_{\pi}\bar{\psi}\gamma^{\mu}\gamma^5\partial_{\mu}\left(\pi^i\tau^i+\eta\right)\psi
-g_{\sigma}\bar{\psi}\sigma\psi\nonumber\\
&&-g_{\rho}\bar{\psi}\gamma^{\mu}\left(\rho_{\mu}^i\tau^i+\omega_{\mu}\right)\psi
-f_{\rho}\bar{\psi}\sigma^{\mu\nu}\partial_{\mu}
\left(\rho_{\nu}^i\tau^i+\omega_{\nu}\right)\psi,\nonumber\\
\end{eqnarray}
where $\tau$ is the flavor spatial operator. Therefore, the OBE effective potentials depend on the spin and isospin of the discussed systems, and have the form of
\begin{eqnarray}
V_{q_1q_2} &=& \mathcal{V}_{\sigma,\eta,\omega}(\bm \sigma_1,\bm \sigma_2)+\bm{\tau_1}\cdot\bm{\tau_2}\mathcal{V}_{\pi,\rho}(\bm \sigma_1,\bm \sigma_2),
\end{eqnarray}
where $q_1$ and $q_2$ are the interacting light quarks, respectively. $\bm{\sigma}_1$ and $\bm{\sigma}_2$ stand for the spin operators. Here, we can simply divide the OBE effective potentials into isospin-related part and isospin-unrelated part, which correspond to the $\pi/\rho$ exchanges and $\sigma/\eta/\omega$ effective potentials, respectively. For the total effective potentials, one should sum over the interactions between all the light quarks in the discussed systems.

Compared to the $\Sigma_c^{(*)}\bar{D}^*$ systems, the OBE effective potentials in the isospin-unrelated part for the $\Xi_c^{(',*)}\bar{D}^*$ systems are very similar as collected in Table \ref{potential}. With the help of the $SU(3)_F$ symmetry, one can further obtain relations for the OBE effective potentials in the isospin-related part between the $\Sigma_c^{(*)}\bar{D}^*$ and $\Xi_c^{(',*)}\bar{D}^*$ systems. The total isospin for the $\Sigma_c^{(*)}\bar{D}^*$ systems is either $I=1/2$ or $I=3/2$, whereas $I=0$ or $I=1$ for the $\Xi_c^{(',*)}\bar{D}^*$ systems. Here, we firstly expand their isospin wave functions $|I_{q_1};I_{q_2},I_{q_3}(I_{q_2q_3});I\rangle$ in terms of the $|I_{q_1},I_{q_2}(I_{q_1q_2});I_{q_3};I\rangle$ basis,
\begin{eqnarray}
|\bar{D}^{(*)}\Sigma_c^{(*)}(I=\frac{1}{2})\rangle &=& \left|\frac{1}{2};\frac{1}{2},\frac{1}{2}(1);\frac{1}{2}\right\rangle\nonumber\\
  &=& \frac{\sqrt{3}}{2}\left|\frac{1}{2},\frac{1}{2}(0);\frac{1}{2};\frac{1}{2}\right\rangle
  +\frac{1}{2}\left|\frac{1}{2},\frac{1}{2}(1);\frac{1}{2};\frac{1}{2}\right\rangle,\nonumber\\\\
|\bar{D}^{(*)}\Sigma_c^{(*)}(I=\frac{3}{2})\rangle &=& \left|\frac{1}{2};\frac{1}{2},\frac{1}{2}(1);\frac{3}{2}\right\rangle
  =\left|\frac{1}{2},\frac{1}{2}(1);\frac{1}{2};\frac{3}{2}\right\rangle,\\
|\bar{D}^{(*)}\Xi_c^{(',*)}(I)\rangle &=& \left|\frac{1}{2};\frac{1}{2},0(\frac{1}{2});I\right\rangle
  =\left|\frac{1}{2},\frac{1}{2}(0);0;I\right\rangle,
\end{eqnarray}
here, we use
\begin{eqnarray}
|I_{q_1};I_{q_2}I_{q_3}(I_{q_2q_3});I\rangle &=&  \sum_{I_{12}}(-1)^{I_1+I_2+I_3+I}\sqrt{(2I_{12}+1)(2I_{23}+1)}\nonumber\\
  &&\times\left\{\begin{array}{ccc}I_1  &I_2  &I_{12}\\I_3  &I &I_{23}\end{array}\right\}|I_1I_2(I_{12});I_3;I\rangle.
\end{eqnarray}
Once sandwiching the isospin operator $\bm{\tau_1}\cdot\bm{\tau_2}$ by the $|I_{q_1},I_{q_2}(I_{q_1q_2});I_{q_3};I\rangle$ basis, we can obtain a serial of relations of the OBE effective potentials in the isospin-related part, i.e.,
\begin{eqnarray}
\mathcal{V}_{\Sigma_c^{(*)}\bar{D}^{(*)}}^{I=1/2} &=& \frac{3}{2}\mathcal{U}^0+\frac{1}{2}\mathcal{U}^1,\label{v1}\\
\mathcal{V}_{\Sigma_c^{(*)}\bar{D}^{(*)}}^{I=3/2} &=& 2\mathcal{U}^1,\\
\mathcal{V}_{\Xi_c^{(',*)}\bar{D}^{(*)}}^{I=0} &=& \mathcal{U}^0,\\
\mathcal{V}_{\Xi_c^{(',*)}\bar{D}^{(*)}}^{I=1} &=& \mathcal{U}^1,\label{v2}
\end{eqnarray}
with
\begin{eqnarray}
\mathcal{U}^0 &=& \left\langle I'_{q_1q_2(q_3)}=0|\bm{\tau_1}\cdot\bm{\tau_2}|I_{q_1q_2(q_3)}=0\right\rangle,\\
\mathcal{U}^1 &=& \left\langle I'_{q_1q_2(q_3)}=1|\bm{\tau_1}\cdot\bm{\tau_2}|I_{q_1q_2(q_3)}=1\right\rangle.
\end{eqnarray}

\renewcommand\tabcolsep{0.12cm}
\renewcommand{\arraystretch}{1.7}
\begin{table}[!htbp]
\caption{The one-boson-exchange effective potentials for the $\Sigma_c^{(*)}\bar{D}^{(*)}$ and $\Xi_c^{(',*)}\bar{D}^{(*)}$ systems. Here, $\sigma=(u\bar u+d\bar d+s\bar s)/\sqrt{3}$, $\eta=(u\bar u+d\bar d-2s\bar s)/\sqrt{6}$, $\omega=(u\bar u+d\bar d)\sqrt{2}$, $\pi^-(\rho^-)=\bar ud$, $\pi^0(\rho^0)=(u\bar u-d \bar d)/\sqrt{2}$, and $\pi^+(\rho^+)=u\bar d$.}\label{potential}
\begin{tabular}{c|ccccc}
\toprule[2pt]
&$\sigma$    &$\eta$  &$\omega$    &$\pi$   &$\rho$\\\hline
$\Sigma_c^{(*)}\bar{D}^{(*)} [I=1/2]$   &$\mathcal{U}_{\sigma}$   &$\mathcal{U}_{\eta}$    &$2\mathcal{U}_{\omega}$
&$\frac{3}{2}\mathcal{U}_{\pi}^0+\frac{1}{2}\mathcal{U}_{\pi}^{1}$     &$\frac{3}{2}\mathcal{U}_{\rho}^0+\frac{1}{2}\mathcal{U}_{\rho}^{1}$ \\
$\Sigma_c^{(*)}\bar{D}^{(*)} [I=3/2]$   &$\mathcal{U}_{\sigma}$   &$\mathcal{U}_{\eta}$    &$2\mathcal{U}_{\omega}$
&$2\mathcal{U}_{\pi}^{1}$     &$2\mathcal{U}_{\rho}^{1}$ \\\hline
$\Xi_c^{(',*)}\bar{D}^{(*)} [I=0]$  &$\mathcal{U}_{\sigma}$   &$-\frac{1}{2}\mathcal{U}_{\eta}$    &$\mathcal{U}_{\omega}$    &$\mathcal{U}_{\pi}^0$     &$\mathcal{U}_{\rho}^0$ \\
$\Xi_c^{(',*)}\bar{D}^{(*)} [I=1]$  &$\mathcal{U}_{\sigma}$   &$-\frac{1}{2}\mathcal{U}_{\eta}$    &$\mathcal{U}_{\omega}$    &$\mathcal{U}_{\pi}^{1}$     &$\mathcal{U}_{\rho}^{1}$ \\
\bottomrule[2pt]
\end{tabular}
\end{table}

In Table \ref{potential}, we summarize the OBE effective potentials for the $\Sigma_c^{(*)}\bar{D}^{(*)}$ and $\Xi_c^{(',*)}\bar{D}^{(*)}$ systems. In Ref. \cite{Chen:2019asm}, we have already prepared the concrete OBE effective potentials for the coupled $\Sigma_c^{(*)}\bar{D}^{(*)}$ systems, and find the interactions from the $\Sigma_c^{(*)}\bar{D}^{(*)}$ systems with $I=1/2$ are strong attractive, for the isoquartet systems, the OBE model provides the weak attractive or repulsive interactions. According to the relations in Table \ref{potential}, we can give a qualitative conclusion that the $S-$wave isoscalar $\Xi_c^{(',*)}\bar{D}^{(*)}$ systems may be the possible strange hidden-charm molecular candidates.

In the heavy quark symmetry, charmed baryons are divided into two multiplets according to the $SU(3)$ flavor symmetry of light quark cluster, $3_F\otimes 3_F=\bar{3}_F\oplus 6_F$, the $\Lambda_c$ and $\Xi_c$ are in the $\bar{3}_F$ multiplet, whereas, the $\Sigma_c^{(*)}$ and $\Xi_c^{(\prime,*)}$ are in the $6_F$ multiplet. Thus, the $\Xi_c\bar{D}^{(*)}$ systems aren't related to the $\Sigma_c^{(*)}\bar{D}^{(*)}$ systems in terms of the $SU(3)$ flavor symmetry and heavy quark symmetry. In the following, we will deduce the OBE effective potentials for the $\Xi_c\bar{D}^{(*)}$ systems. The general procedures of the derivations of the OBE effective potentials include three steps. After constructed the effective Lagrangians, one can firstly write down the scattering amplitudes $\mathcal{M}_{\text{OBE}}(h_1h_2\to h_3h_4)$ for the discussed processes. Then, the OBE effective potentials $\mathcal{V}(\bm{q})$ can be related to the corresponding scattering amplitudes by using the Breit approximation, $\mathcal{V}(\bm{q}) = -\mathcal{M}(h_1h_2\to h_3h_4)/\sqrt{\prod_i2M_i\prod_f2M_f}$ with $M_i$ and $M_f$ being the masses of the initial states ($h_1$, $h_2$) and final states ($h_3$, $h_4$), respectively. At last, one can obtain the OBE effective potentials in the coordinate space $\mathcal{V}(\bm{r})$ after performing the Fourier transformation, i.e.,
\begin{eqnarray}
\mathcal{V}_{E}^{h_1h_2\to h_3h_4}(\bm{r}) =
          \int\frac{d^3\bm{q}}{(2\pi)^3}e^{i\bm{q}\cdot\bm{r}}
          \mathcal{V}_{E}^{h_1h_2\to h_3h_4}(\bm{q})\mathcal{F}^2(q^2,m_E^2).\nonumber
\end{eqnarray}
Here, for compensating the off-shell effect of the exchanged bosons, a monopole form factor $\mathcal{F}(q^2,m_E^2)= (\Lambda^2-m_E^2)/(\Lambda^2-q^2)$ is introduced at every interactive vertex, where $m_E$ and $q$ are the mass and four-momentum of the exchanged meson, respectively. $\Lambda$ is the cutoff. A reasonable cutoff value is around 1.00 GeV \cite{Tornqvist:1993ng,Tornqvist:1993vu}. Especially, we also take this empirical value in reproducing the masses of the $P_c(4312)$, $P_c(4440)$, and $P_c(4457)$\cite{Chen:2019asm}.

The relevant effective Lagrangians are constructed in the heavy quark limit and chiral symmetry \cite{Yan:1992gz,Wise:1992hn,Burdman:1992gh,Casalbuoni:1996pg,Falk:1992cx,Liu:2011xc}, i.e.,
\begin{eqnarray}
\mathcal{L}_{H}&=&g_S\langle \bar{H}^{(\overline{Q})}_a\sigma H^{(\overline{Q})}_b\rangle+ig\langle \bar{H}^{(\overline{Q})}_a\gamma_{\mu}{\mathcal A}_{ab}^{\mu}\gamma_5H^{(\overline{Q})}_b\rangle\nonumber\\
  &&-i\beta\langle \bar{H}^{(\overline{Q})}_av_{\mu}\left(\mathcal{V}^{\mu}-\rho^{\mu}\right)_{ab}H^{(\overline{Q})}_b\rangle\nonumber\\
  &&+i\lambda\langle \bar{H}^{(\overline{Q})}_a\sigma_{\mu\nu}F^{\mu\nu}(\rho)H^{(\overline{Q})}_b\rangle,\label{lag1}\\
\mathcal{L}_{\mathcal{B}_{\bar{3}}} &=& l_B\langle\bar{\mathcal{B}}_{\bar{3}}\sigma\mathcal{B}_{\bar{3}}\rangle +i\beta_B\langle\bar{\mathcal{B}}_{\bar{3}}v^{\mu}(\mathcal{V}_{\mu}-\rho_{\mu})\mathcal{B}_{\bar{3}}\rangle,\nonumber\\
\mathcal{L}_{\mathcal{B}^{(',*)}_6} &=&  l_S\langle\bar{\mathcal{S}}_{\mu}\sigma\mathcal{S}^{\mu}\rangle
         -\frac{3}{2}g_1\varepsilon^{\mu\nu\lambda\kappa}v_{\kappa}\langle\bar{\mathcal{S}}_{\mu}{\mathcal A}_{\nu}\mathcal{S}_{\lambda}\rangle\nonumber\\
  &&+i\beta_{S}\langle\bar{\mathcal{S}}_{\mu}v_{\alpha}\left(\mathcal{V}^{\alpha}-\rho^{\alpha}\right) \mathcal{S}^{\mu}\rangle
         +\lambda_S\langle\bar{\mathcal{S}}_{\mu}F^{\mu\nu}(\rho)\mathcal{S}_{\nu}\rangle,\nonumber\\
\mathcal{L}_{\mathcal{B}_{\bar{3}}\mathcal{B}^{(',*)}_6} &=& ig_4\langle\bar{\mathcal{S}^{\mu}}{\mathcal A}_{\mu}\mathcal{B}_{\bar{3}}\rangle
         +i\lambda_I\varepsilon^{\mu\nu\lambda\kappa}v_{\mu}\langle \bar{\mathcal{S}}_{\nu}F_{\lambda\kappa}\mathcal{B}_{\bar{3}}\rangle+h.c..\nonumber\\\label{lag2}
\end{eqnarray}
In Eqs. (\ref{lag1})-(\ref{lag2}), the multiplet fields $H^{(\bar{Q})}$ and $\mathcal{S}$ are linear combinations of the $S-$wave charmed mesons and charmed baryons in the $6_F$ flavor representation, respectively. $H^{(\bar{Q})} = [\tilde{\mathcal{P}}^{*\mu}\gamma_{\mu}-\tilde{\mathcal{P}}\gamma_5]\frac{1-\rlap\slash v}{2}$ with $\tilde{\mathcal{P}}=\left(\bar{D}^0,\,D^-\right)^T$ and $\tilde{\mathcal{P}}^*=\left(\bar{D}^{*0},\,D^{*-}\right)^T$. $\mathcal{S}_{\mu} =
-\sqrt{\frac{1}{3}}(\gamma_{\mu}+v_{\mu})\gamma^5\mathcal{B}_6^{(\prime)}+\mathcal{B}_{6\mu}^*$. $A_{\mu}$ and $\mathcal{V}_{\mu}$ correspond to the axial current and vector current, respectively, $A_{\mu} = \frac{1}{2}(\xi^{\dag}\partial_{\mu}\xi-\xi\partial_{\mu}\xi^{\dag})=\frac{i}{f_{\pi}}
\partial_{\mu}\mathbb{P}+\ldots$, and $\mathcal{V}_{\mu} =
\frac{1}{2}(\xi^{\dag}\partial_{\mu}\xi-\xi\partial_{\mu}\xi^{\dag})
=\frac{i}{2f_{\pi}^2}\left[\mathbb{P},\partial_{\mu}\mathbb{P}\right]+\ldots$ with $\xi=\text{exp}(i\mathbb{P}/f_{\pi})$ and the pion decay constant $f_{\pi}=132$ MeV. $\rho_{ba}^{\mu}=ig_V\mathbb{V}_{ba}^{\mu}/\sqrt{2}$, $F^{\mu\nu}(\rho)=\partial^{\mu}\rho^{\nu}-\partial^{\nu}\rho^{\mu}
+\left[\rho^{\mu},\rho^{\nu}\right]$. $\mathbb{P}$ and $\mathbb{V}$ stand for the isoscalar and vector matrixes, respectively. Matrices for the $\mathcal{B}_{\bar{3}}$, $\mathcal{B}_6^{(',*)}$, $\mathbb{P}$, and $\mathbb{V}$ are expressed as
\begin{eqnarray*}
\mathcal{B}_{\bar{3}} &=& {\left(\begin{array}{ccc}
         0    &\Lambda_c^+       &\Xi_c^+\\
        -\Lambda_c^+      &0     &\Xi_c^0\\
       -\Xi_c^+      &-\Xi_c^0    &0
                \end{array}\right),}\\
\mathcal{B}_6^{(',*)} &=& {\left(\begin{array}{ccc}
         \Sigma_c^{(*)++}              &\frac{1}{\sqrt{2}}\Sigma_c^{(*)+}    &\frac{1}{\sqrt{2}}\Xi_c^{(',*)+}\\
         \frac{1}{\sqrt{2}}\Sigma_c^{(*)+}      &\Sigma_c^{(*)0}       &\frac{1}{\sqrt{2}}\Xi_c^{(',*)0}\\
          \frac{1}{\sqrt{2}}\Xi_c^{(',*)+}     &\frac{1}{\sqrt{2}}\Xi_c^{(',*)0}      &\Omega_c^{(*)0}\\
                \end{array}\right)},\\
\mathbb{P} &=& {\left(\begin{array}{ccc}
       \frac{\pi^0}{\sqrt{2}}+\frac{\eta}{\sqrt{6}} &\pi^+      &K^+\\
       \pi^-       &-\frac{\pi^0}{\sqrt{2}}+\frac{\eta}{\sqrt{6}}     &K^0\\
       K^-         &\bar{K}^0      &-\frac{2}{\sqrt{6}}\eta
               \end{array}\right)},\\
{V} &=& {\left(\begin{array}{ccc}
\frac{\rho^0}{\sqrt{2}}+\frac{\omega}{\sqrt{2}}  &\rho^+      &K^{*+}\\
\rho^- &-\frac{\rho^0}{\sqrt{2}}+\frac{\omega}{\sqrt{2}}      &K^{*0}\\
K^{*-}      &\bar{K}^{*0}    &\phi
\end{array}\right)},
\end{eqnarray*}
respectively.

In this work, we adopt the same values of the coupling constants in Ref. \cite{Liu:2011xc,Chen:2019asm,Chen:2020kco}, where $g=0.59$, $g_1=0.94$, and $g_4=1.06$ are extracted from the decay widths of $\Gamma(D^*\to D\pi)$ and $\Gamma(\Sigma_c^{(*)}\to \Lambda_c\pi)$ \cite{Isola:2003fh,Liu:2011xc,pdg}, respectively. In the charmed mesons sector, $g_S=0.76$ \cite{Bardeen:2003kt}, $\beta =0.9$ \cite{Isola:2003fh}, $\lambda =0.56$ GeV$^{-1}$ \cite{Isola:2003fh}, $g_V=m_{\rho}/f_{\pi}=5.9$. For the remaining coupling constants between the heavy baryons and the light meson ($\sigma$, $\rho$, $\omega$), their values are estimated from the nucleon-nucleon interactions \cite{Liu:2011xc}, $l_B=-3.65$, $\beta_Bg_V=6.0$, $l_S=6.2$, $\beta_Sg_V=-12.0$, $\lambda_Sg_V=-19.2$ $\text{GeV}^{-1}$, $\lambda_Ig_V=6.8$ $\text{GeV}^{-1}$. With these preparations, we can deduce the detailed expressions of the OBE effective potentials, i.e.,
\begin{eqnarray}
\mathcal{V}^{\Xi_c\bar{D}\to\Xi_c\bar{D}}(r) &=&
    2\mathbb{A}Y_{\Lambda,m_{\sigma}}
    +\frac{\mathbb{B}}{4}\left(\mathcal{G}(I)Y_{\Lambda,m_{\rho}}+Y_{\Lambda,m_{\omega}}\right),\quad\label{pot1}\\
\mathcal{V}^{\Xi_c^{\prime}\bar{D}\to\Xi_c\bar{D}}(r) &=& 0,\\
\mathcal{V}^{\Xi_c\bar{D}^*\to\Xi_c\bar{D}}(r) &=& 0,\\
\mathcal{V}^{\Xi_c^*\bar{D}\to\Xi_c\bar{D}}(r) &=& 0,\\
\mathcal{V}^{\Xi_c^{\prime}\bar{D}^*\to\Xi_c\bar{D}}(r) &=&
    \frac{\mathbb{C}}{6\sqrt{6}}\left(\mathcal{G}(I)\mathcal{Z}^{13}_{\Lambda_1,m_{\pi1}}
    +\mathcal{Z}^{13}_{\Lambda_1,m_{\eta1}}\right)\nonumber\\
    &&-\frac{\mathbb{D}}{3\sqrt{6}}\left(\mathcal{G}(I)\mathcal{Z}^{\prime13}_{\Lambda_1,m_{\rho1}}
   +\mathcal{Z}^{\prime13}_{\Lambda_1,m_{\omega1}}\right),\\
\mathcal{V}^{\Xi_c^*\bar{D}^*\to\Xi_c\bar{D}}(r) &=&
    -\frac{\mathbb{C}}{6\sqrt{2}}\left(\mathcal{G}(I)\mathcal{Z}^{14}_{\Lambda_2,m_{\pi2}}
    +\mathcal{Z}^{14}_{\Lambda_2,m_{\eta2}}\right)\nonumber\\
    &&-\frac{\mathbb{D}}{6\sqrt{2}}\left(\mathcal{G}(I)\mathcal{Z}^{\prime14}_{\Lambda_2,m_{\rho2}}
   +\mathcal{Z}^{\prime14}_{\Lambda_2,m_{\omega2}}\right),\label{pot2}
\end{eqnarray}
for the $\Xi_c^{(\prime,*)}\bar{D}^{(*)}\to\Xi_c\bar{D}$ processes, where $\mathbb{A}=l_Bg_s$, $\mathbb{B}=\beta\beta_Bg_v^2$, $\mathbb{C}=gg_4/f_{\pi}^2$, $\mathbb{D}=\lambda\lambda_Ig_v^2$, $\mathcal{G}(I)$ is the isospin factor, which is taken as $1$ for the isospin-$1$ system, and $-3$ for the isospin-$0$ system. The functions $Y_{\Lambda,m}$, $\mathcal{Z}^{ij}_{\Lambda,m_a}$, and $\mathcal{Z}^{\prime ij}_{\Lambda,m_a}$ denote
\begin{eqnarray}
Y_{\Lambda,m} &=&\frac{1}{4\pi r}(e^{-mr}-e^{-\Lambda r})-\frac{\Lambda^2-m^2}{8\pi \Lambda}e^{-\Lambda r},\\
\mathcal{Z}^{ij}_{\Lambda, m_a} &=&\left(\mathcal{E}_{ij}\nabla^2+\mathcal{F}_{ij}r\frac{\partial}{\partial r}\frac{1}{r}\frac{\partial}{\partial r}\right)Y_{\Lambda,m_a},\\
\mathcal{Z}^{\prime ij}_{\Lambda,m_a}&=&\left(2\mathcal{E}_{ij}\nabla^2-\mathcal{F}_{ij}r\frac{\partial}{\partial
r}\frac{1}{r}\frac{\partial}{\partial r}\right)Y_{\Lambda,m_a},
\end{eqnarray}
The variables in Eqs. (\ref{pot1})-(\ref{pot2}) are defined as $\Lambda_i^2 =\Lambda^2-q_i^2$, $m_{{i}}^2=m^2-q_i^2$, with $i=1,2$. $q_1=0.64$ MeV and $q_2=38.14$ MeV. The spin-spin interaction and tensor force operators read as $\mathcal{E}_{13}=\vec{\sigma}\cdot\vec{\epsilon}_4^{\dag}$, $\mathcal{E}_{14}=\sum_{1/2,a;1,b}^{3/2,a+b}\chi_{3,a}^{\dag}\vec{\epsilon}_{3,b}^{\dag}\cdot\vec{\epsilon}_4^{\dag}\chi_1$, $\mathcal{F}_{13}=S\left(\hat{r},\vec{\sigma},\vec{\epsilon}_4^{\dag}\right)$, and $\mathcal{F}_{14}=\sum_{1/2,a;1,b}^{3/2,a+b}\chi_{3,a}^{\dag}S\left(\hat{r},\vec{\epsilon}_{3,b}^{\dag},\vec{\epsilon}_4^{\dag}\right)\chi_1$. In addition, one can refer to the concrete subpotentials for the $\Xi_c^{(\prime,*)}\bar{D}^{(*)}\to\Xi_c\bar{D}^*$ processes in Ref. \cite{Chen:2020kco}.

\section{Numerical results}\label{sec3}


Before producing numerical calculations, we would like to make several remarks on the bound state solutions for the reasonable loosely bound molecular state after performing the coupled channel analysis \cite{Chen:2017xat},
\begin{enumerate}
  \item The reasonable cutoff value in the monopole form factor is around 1.00 GeV according to the experience of the nucleon-nucleon interaction \cite{Tornqvist:1993ng,Tornqvist:1993vu}.
  \item The binding energy is around several to a few tens MeV.
  \item The root-mean-square (RMS) radii are around a few fm or larger as the size of the loosely bound molecule should be much larger than the size of the component hadrons.
  \item For an $S-$wave molecular state composed by two mesons, the asymptotic form of its wave function can be expressed as $\psi(r)\sim e^{-\sqrt{2\mu E}r}/r$, where $\mu$ and $E$ stand for the reduced mass and the binding energy, respectively. In the coupled channel analysis, the $\mu$ is the reduced mass for the dominant channel, and $E$ is measured from the dominant channel, $E=M_{\text{lowest}}-M_{\text{dominant}}+E_{\text{binding}}$. When we use the approximated wave function, we can obtain the relation between the molecular size and its binding energy, $R\sim 1/\sqrt{2\mu E}$ \cite{Chen:2017xat,Close:2003sg}. By using this relation, the system with the lowest mass threshold is the dominant channel.
\end{enumerate}

With the above preparations, we firstly solve the coupled channel Shr$\ddot{\text{o}}$dinger equation to find the bound state solutions (including the binding energies $E$, the RMS radii $r_{RMS}$, and the probabilities for all the discussed channels $p_i$) for the $S-$wave coupled $\Xi_c^{(',*)}\bar{D}^{(*)}$ systems with all the possible quantum numbers. The cutoff value is taken from 0.8 GeV to 5.0 GeV. And then, we check whether the obtained bound state solutions match the above criterion of the reasonable loosely bound molecular states (the binding energy around a few to 10 MeV and the RMS radius around or larger than 1 fm). Finally, we compare their cutoff value in these reasonable loosely bound states. In our analysis, we conclude that the state with the obtained reasonable loosely bound state solutions in the cutoff range $\Lambda$ around 1.00 GeV is the prime loosely bound molecular candidates. With the increasing of the cutoff value, the possible of the existence of the loosely bound molecule becomes lower. For the reasonable loosely bound states with the cutoff $1.00<\Lambda<2.00$ GeV, they may be the possible molecular candidate. If the cutoff $\Lambda$ is larger than 3.00 GeV, we conclude that they cannot be a good molecular candidate.

\begin{figure}[!htbp]
\center
\includegraphics[width=1.6in]{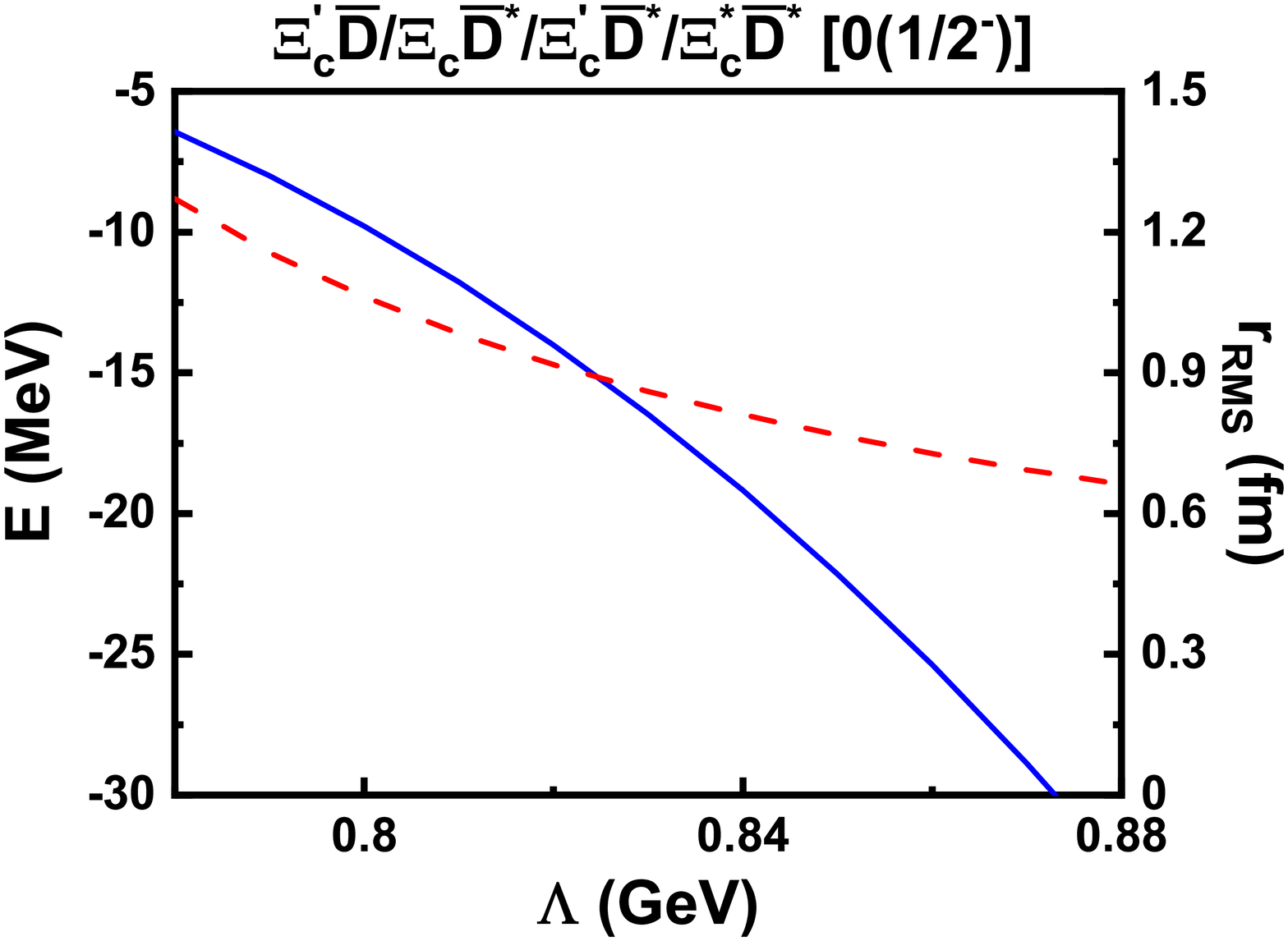}
\includegraphics[width=1.6in]{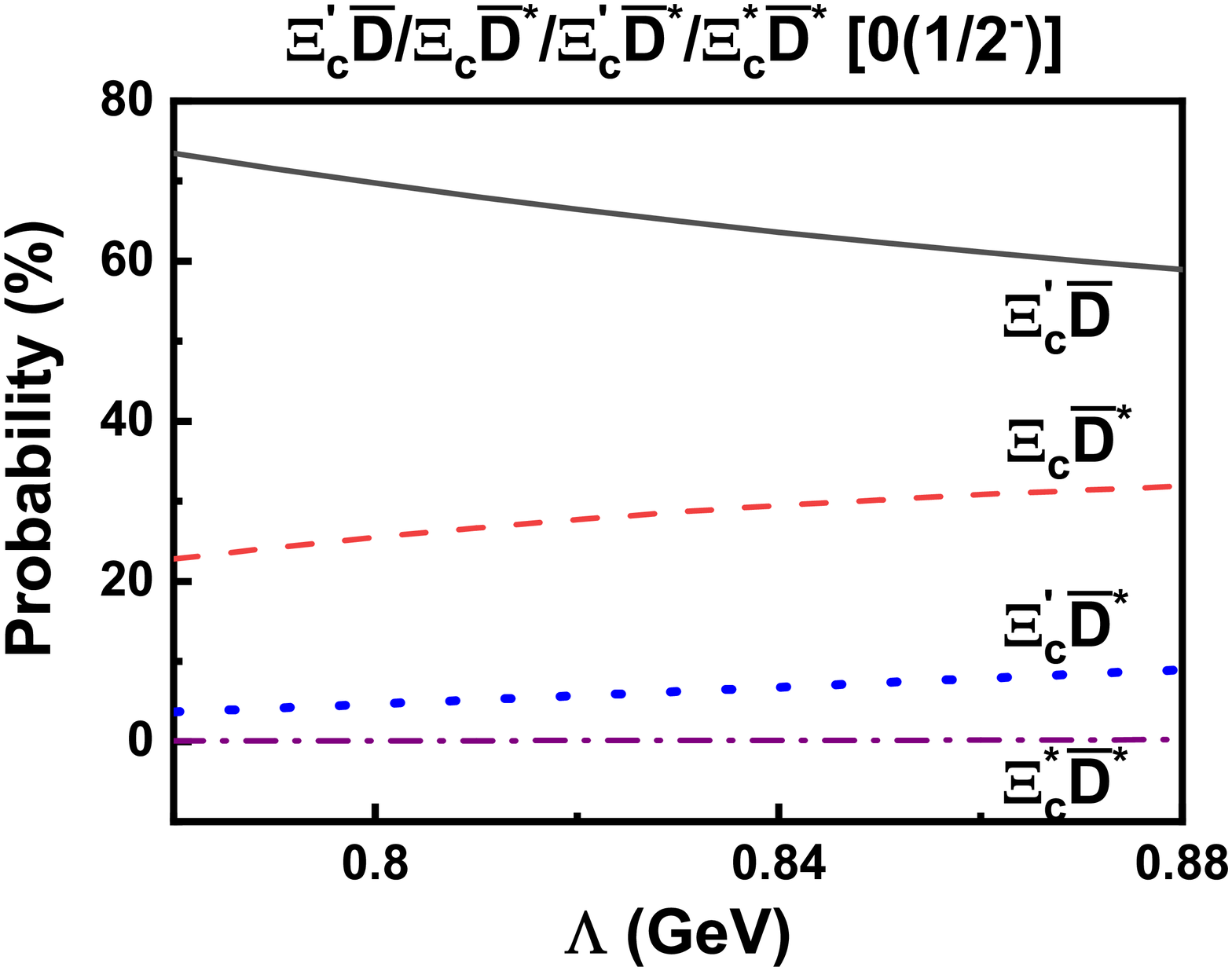}\\
\includegraphics[width=1.6in]{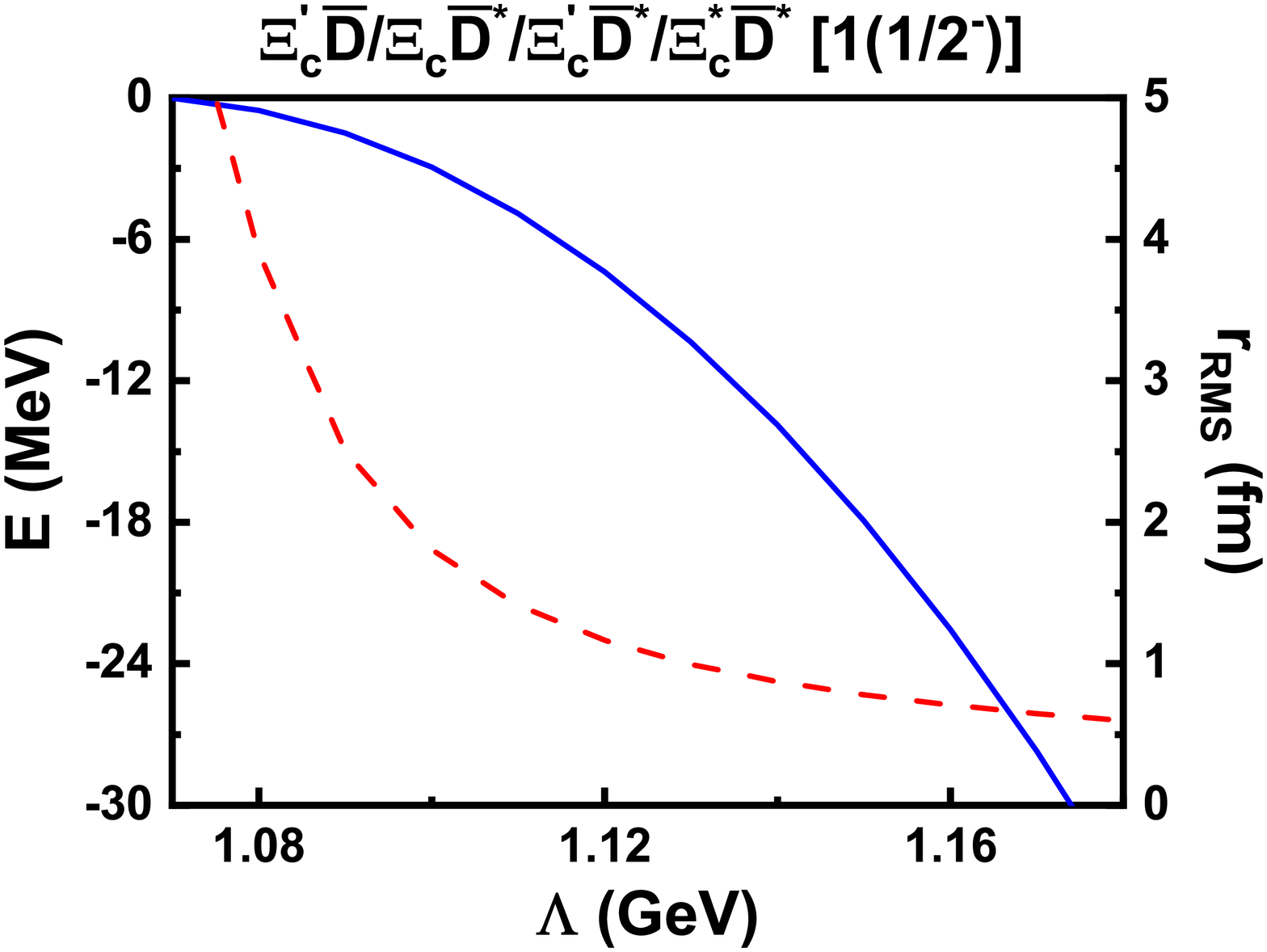}
\includegraphics[width=1.6in]{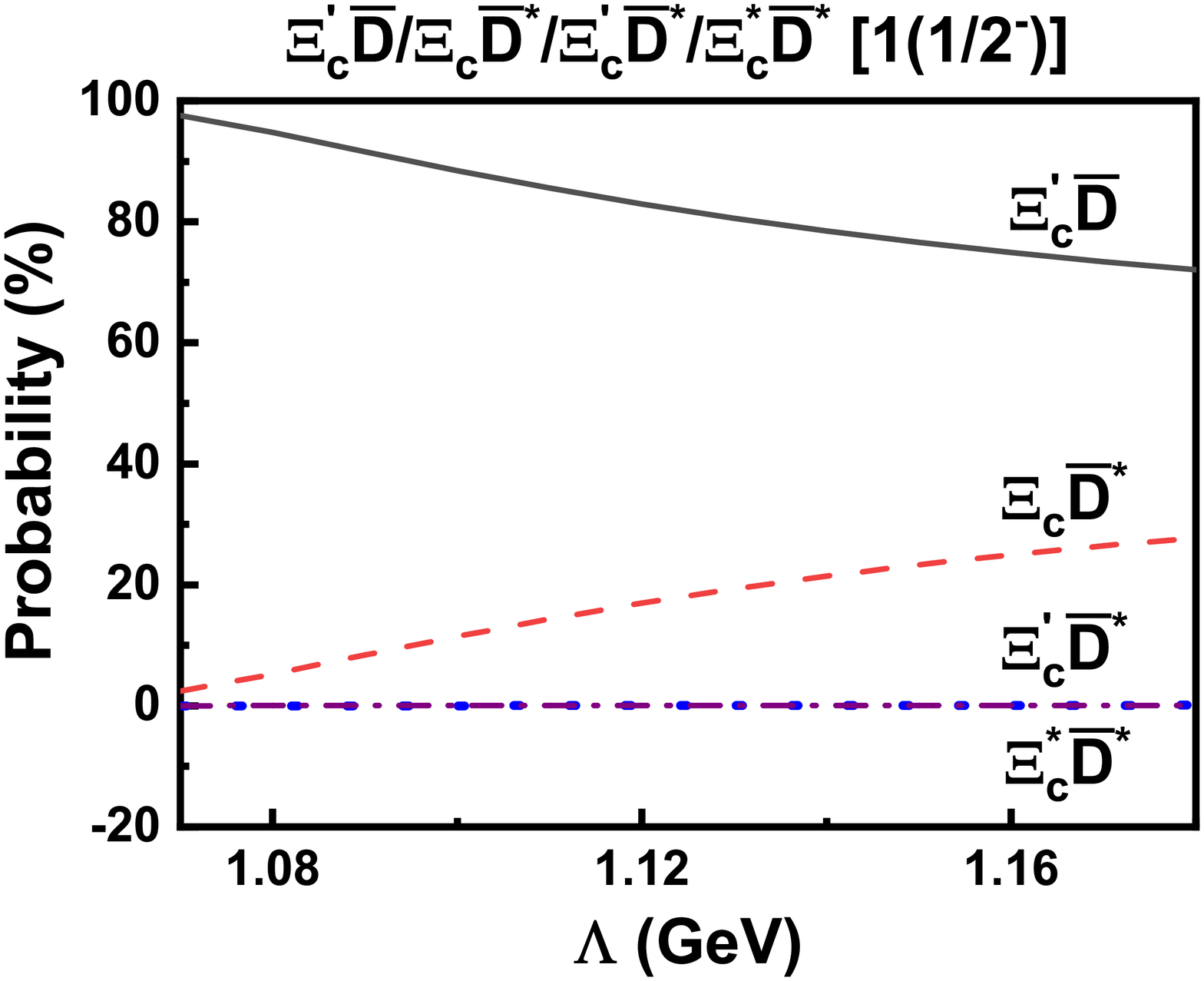}\\
\includegraphics[width=1.6in]{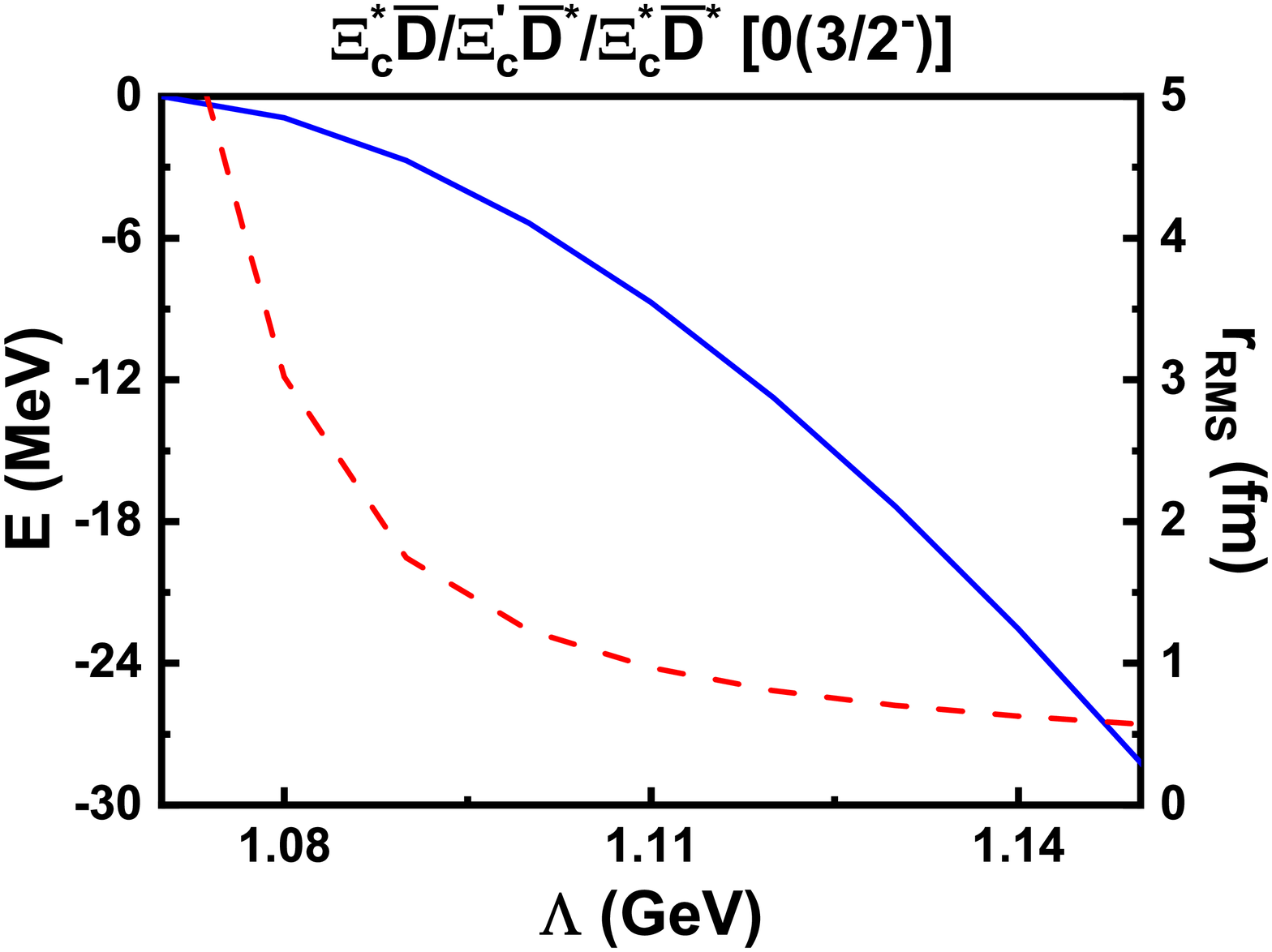}
\includegraphics[width=1.6in]{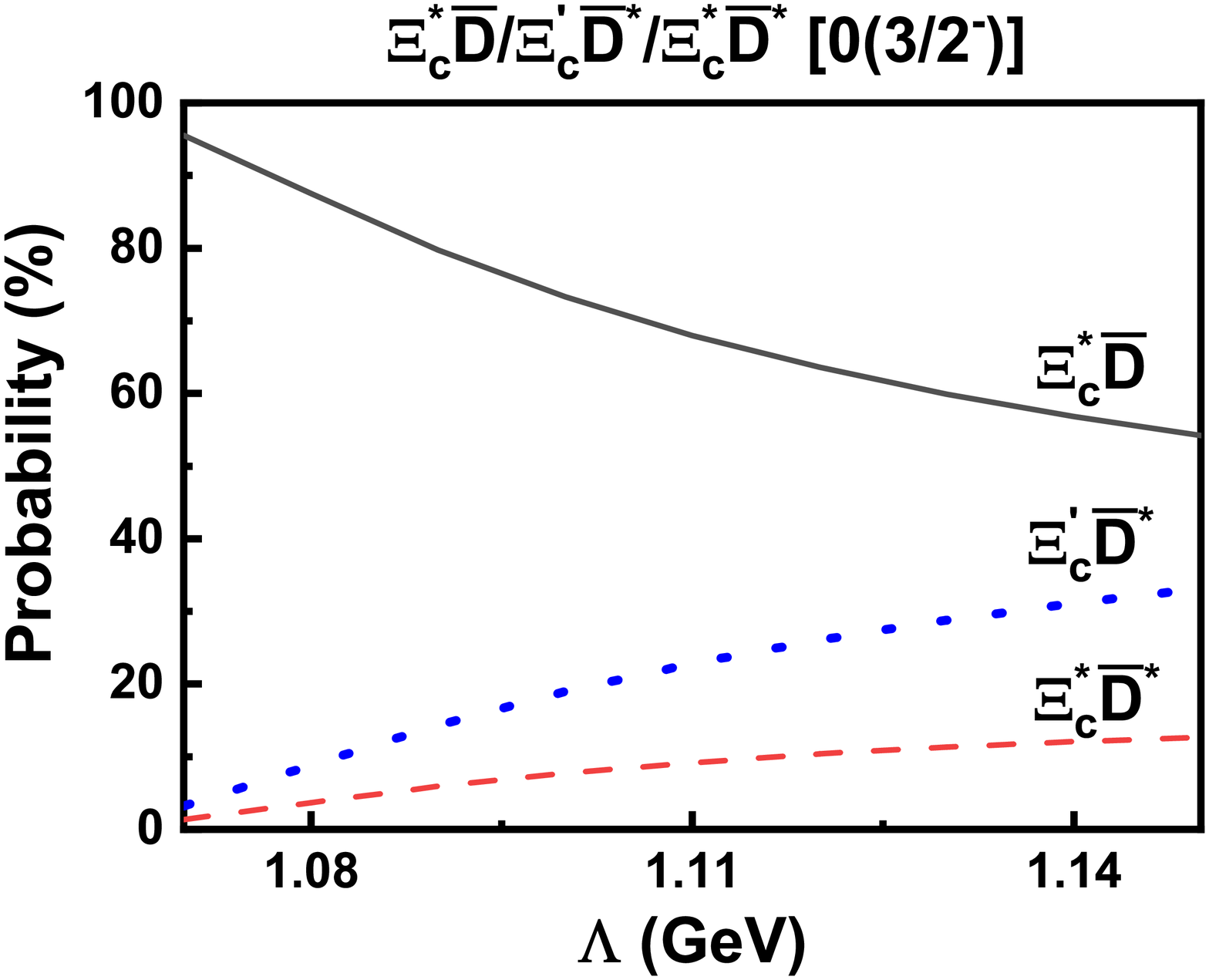}\\
\includegraphics[width=1.6in]{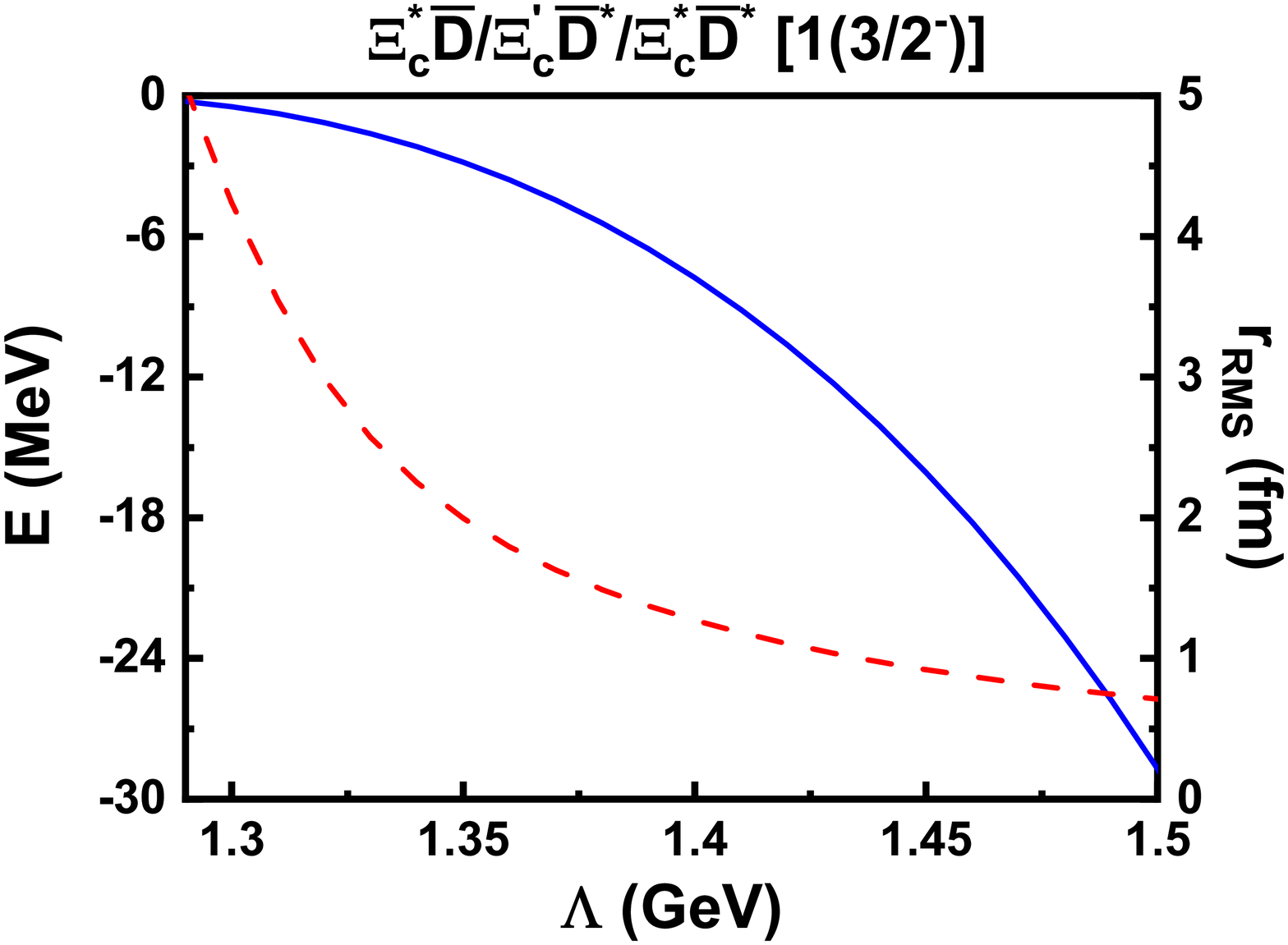}
\includegraphics[width=1.6in]{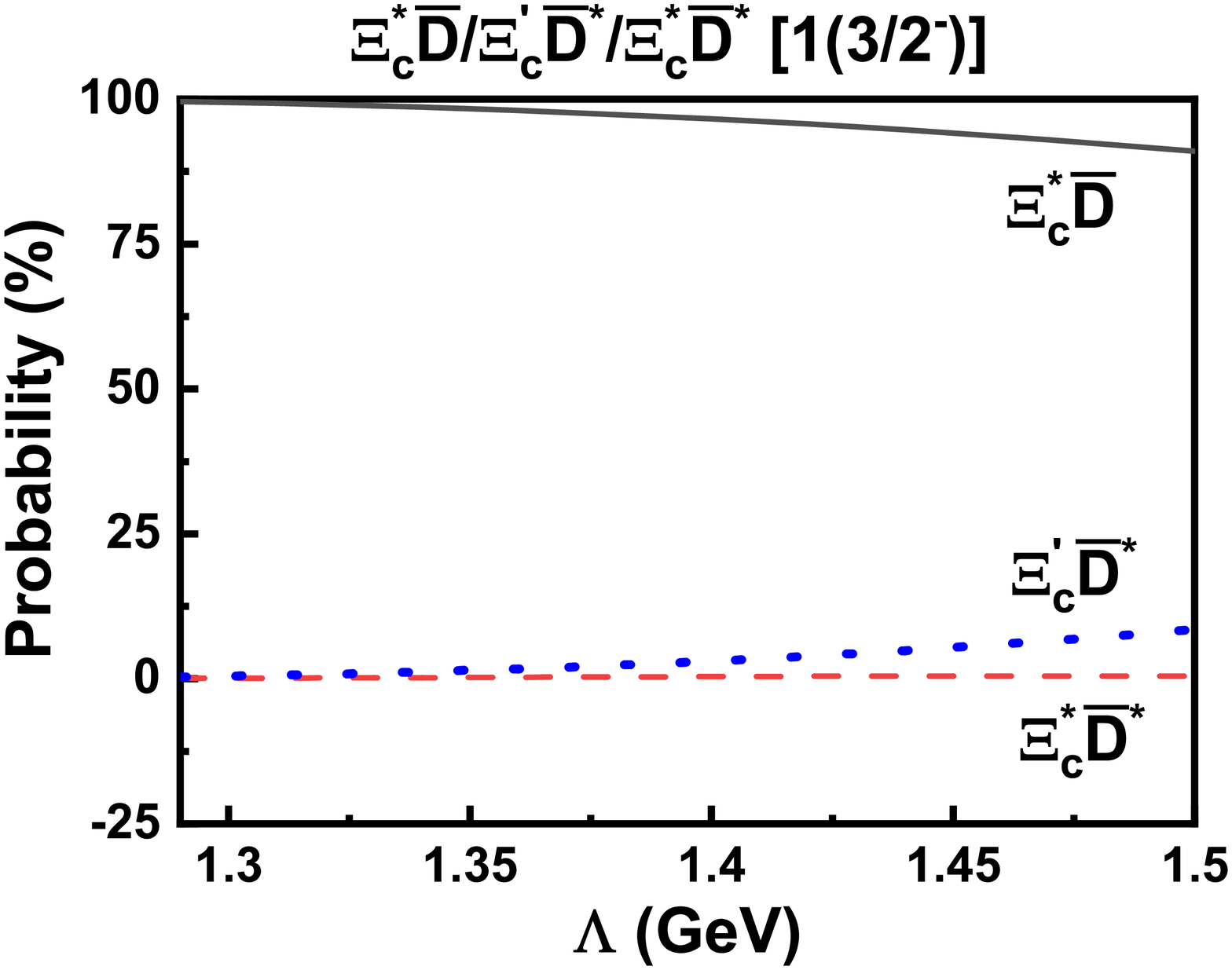}\\
\includegraphics[width=1.6in]{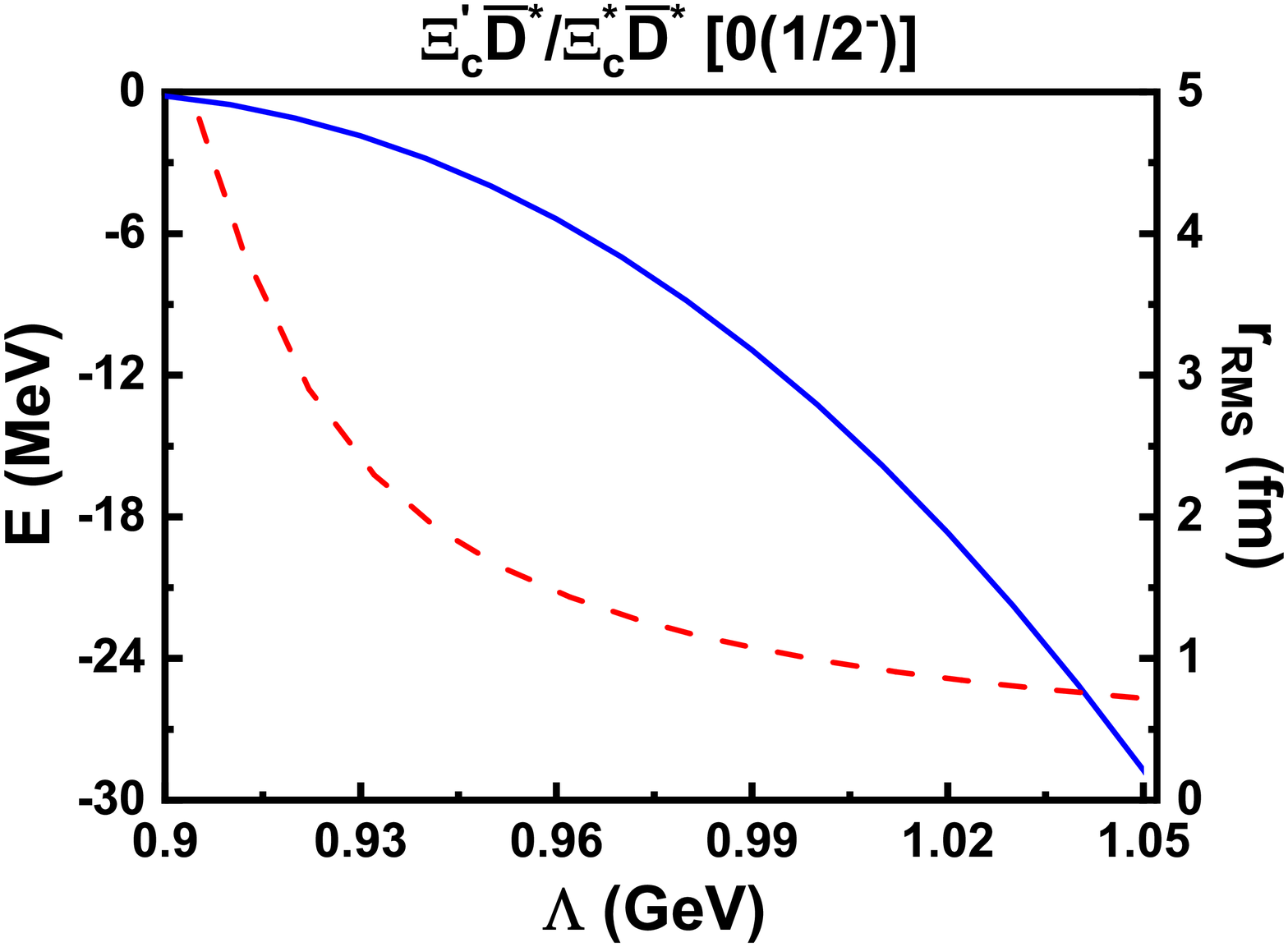}
\includegraphics[width=1.6in]{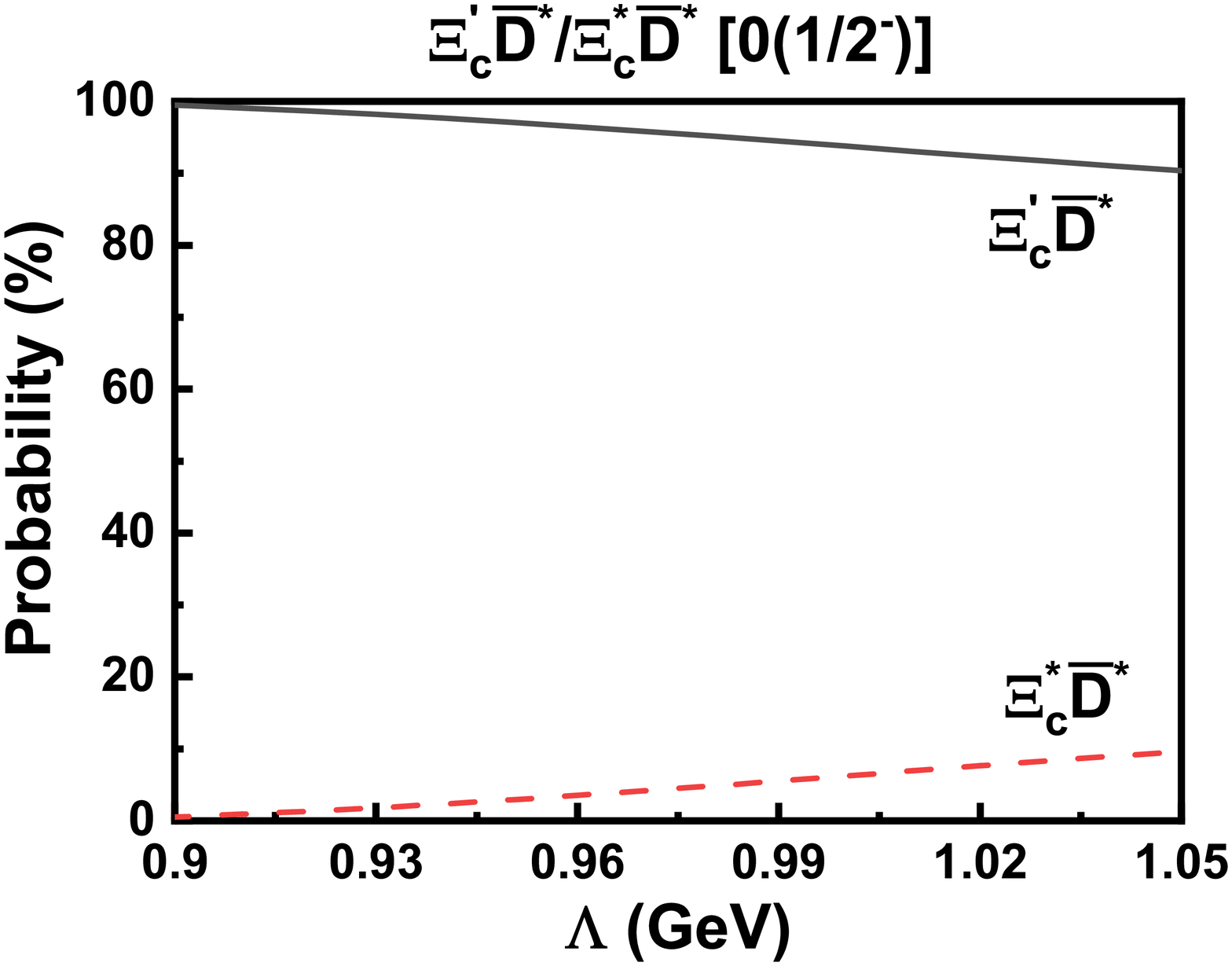}\\
\includegraphics[width=1.6in]{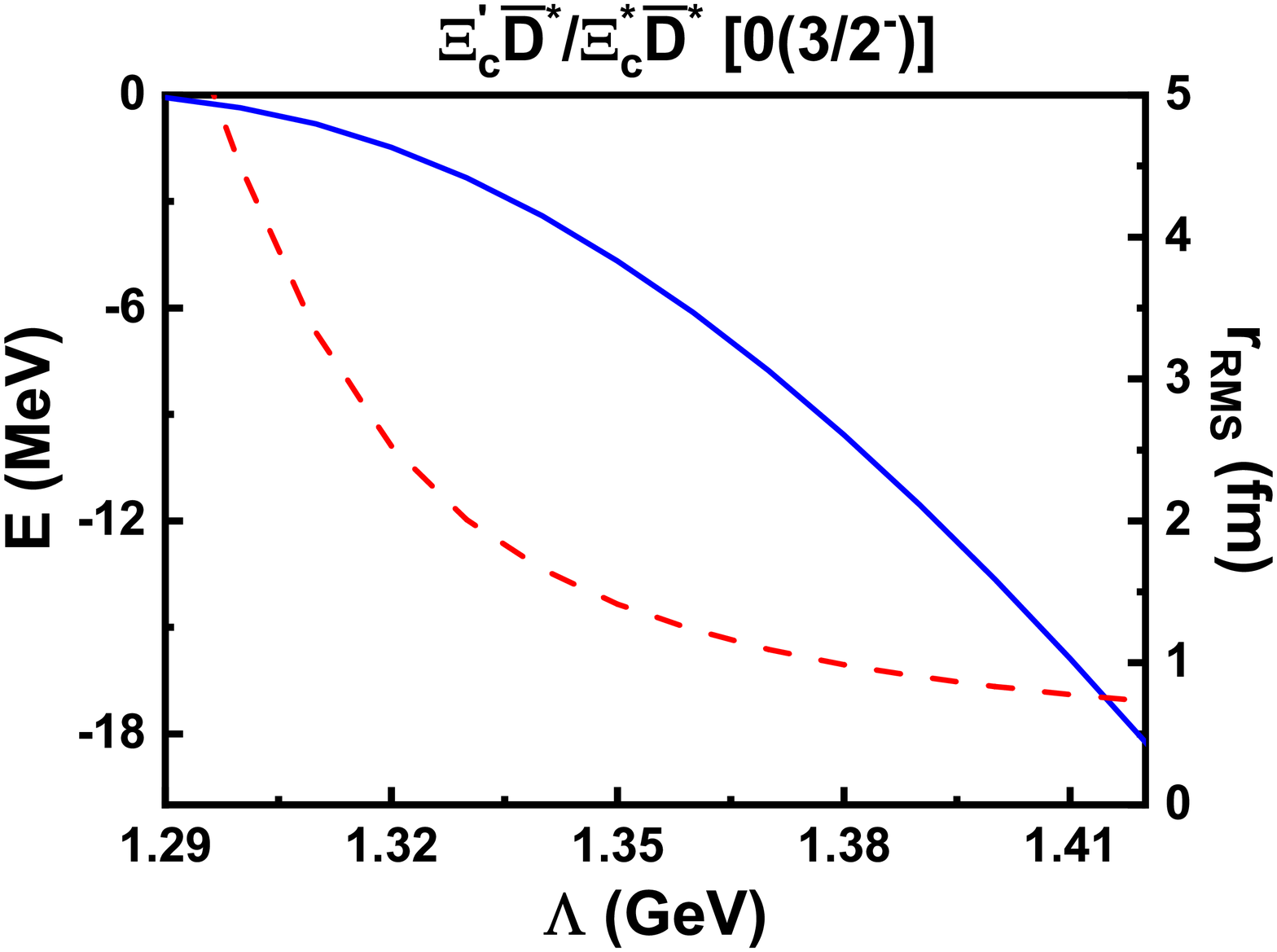}
\includegraphics[width=1.6in]{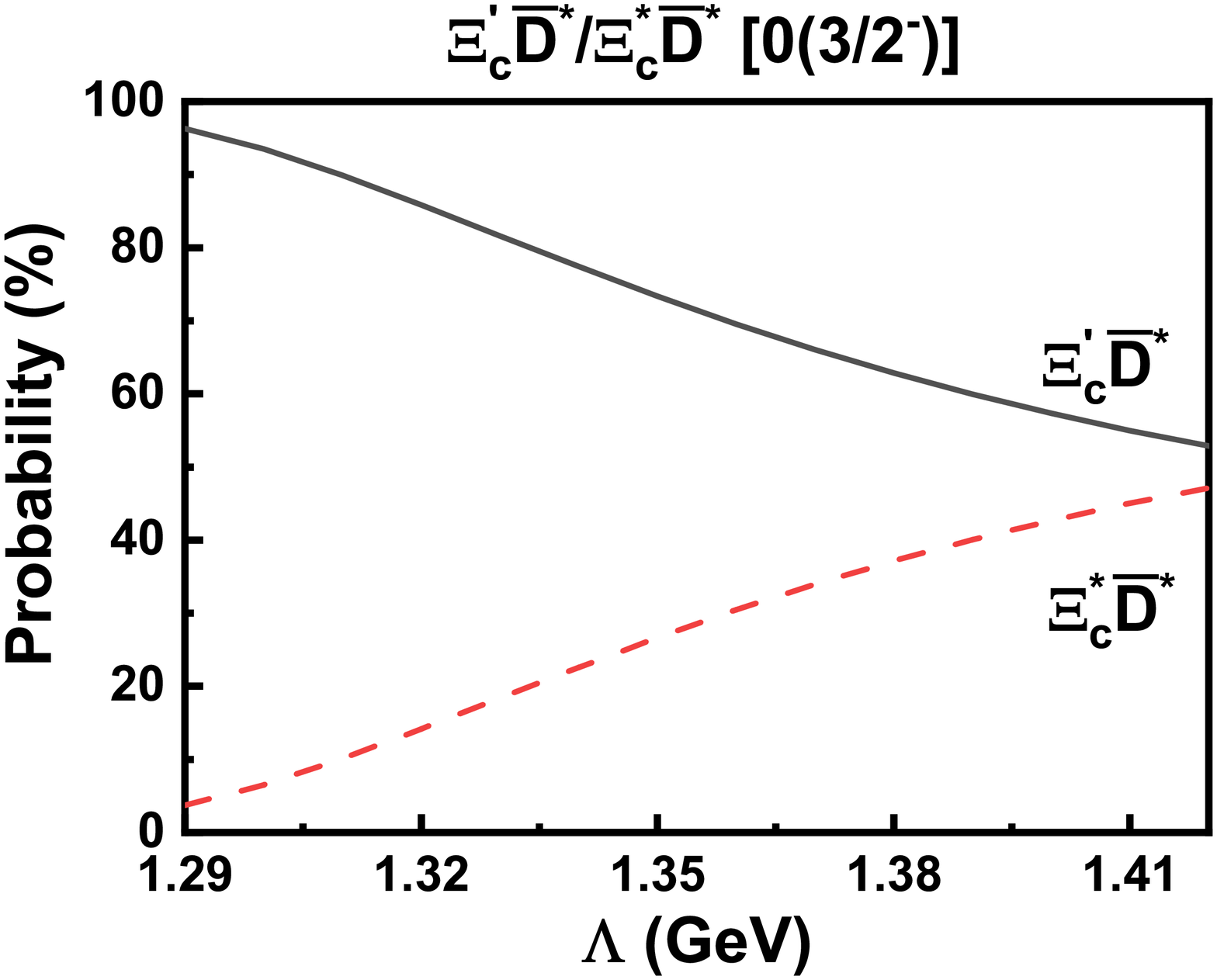}\\
\includegraphics[width=1.6in]{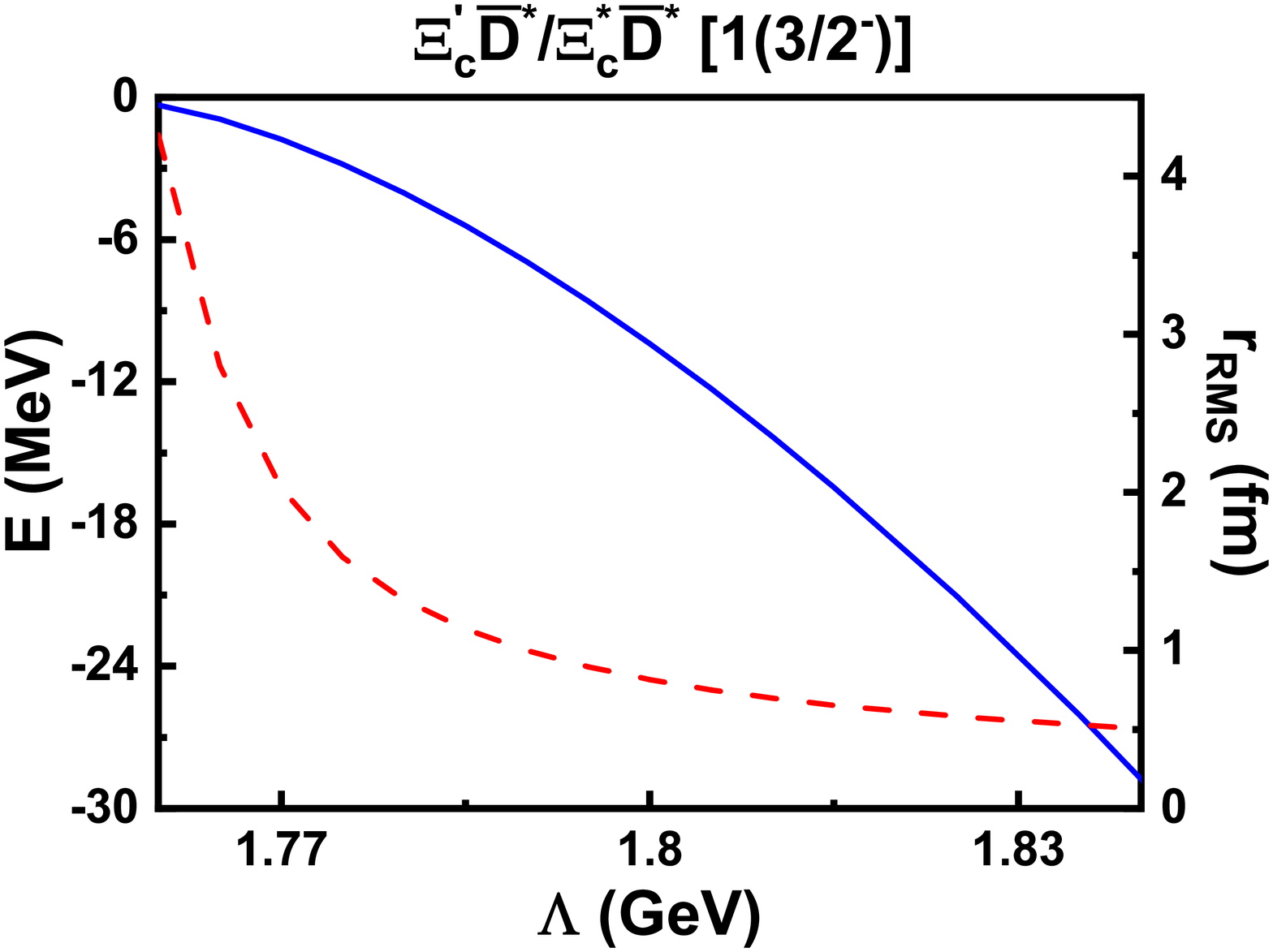}
\includegraphics[width=1.6in]{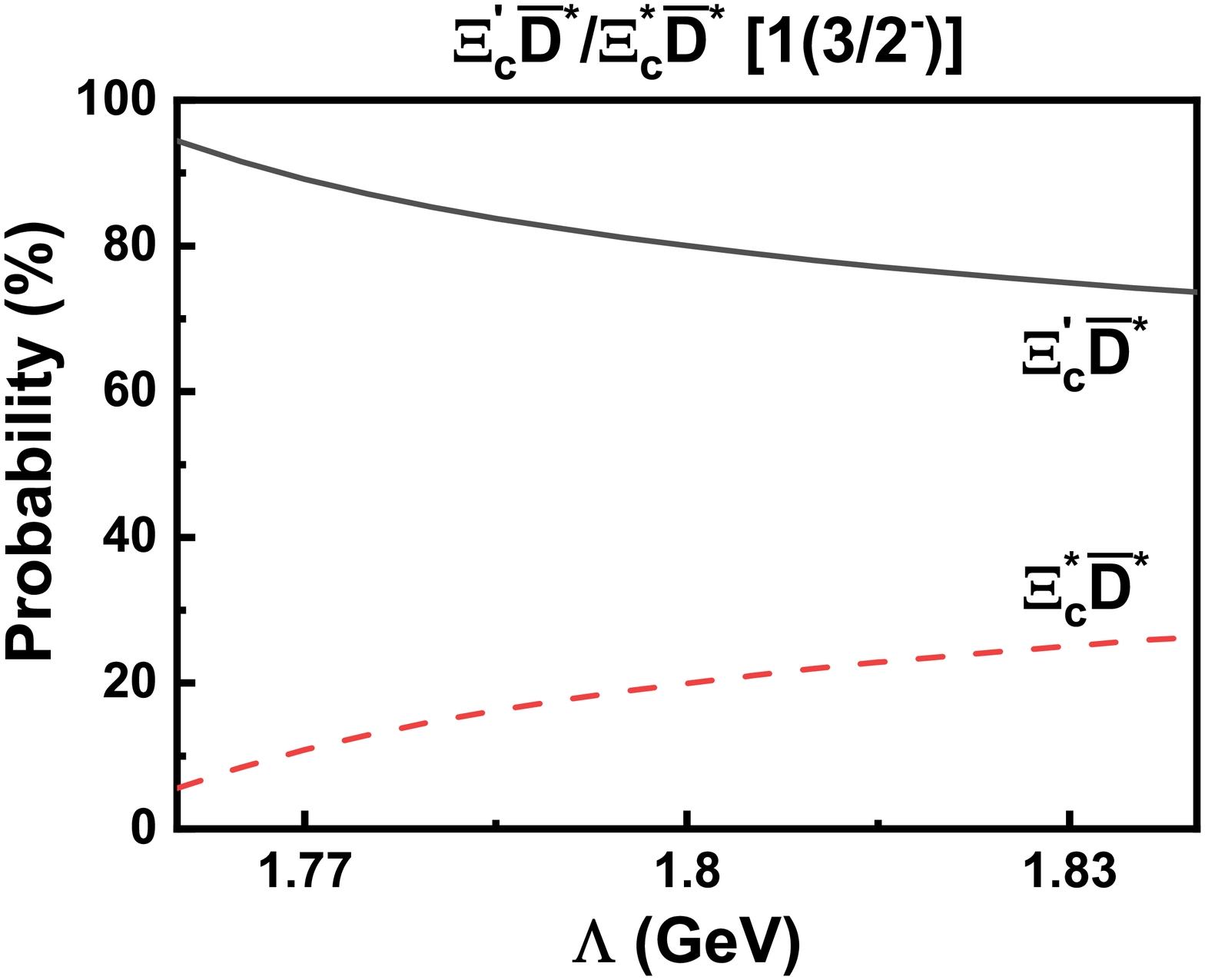}
\caption{The $\Lambda$ dependence of the binding energy $E$, root-mean-square radii, and the probabilities for all discussed channels. The blue solid lines and the red dashed lines in the left figures correspond to the binding energies curves and root-mean-square radii curves, respectively.}\label{loose}
\end{figure}

As shown in Figure \ref{loose}, we present the $\Lambda$ dependence of the bound state properties ($E$, $r_{RMS}$, and $p_i$) for the coupled $\Xi_c^{\prime}\bar{D}/\Xi_c\bar{D}^*/\Xi_c^{\prime}\bar{D}^*/\Xi_c^*\bar{D}^*$ states with $0,1(1/2^-)$, the coupled $\Xi_c^{*}\bar{D}/\Xi_c^{\prime}\bar{D}^*/\Xi_c^*\bar{D}^*$ states with $0,1(3/2^-)$, and the coupled $\Xi_c^{\prime}\bar{D}^*/\Xi_c^*\bar{D}^*$ states with $0(1/2^-)$ and $0,1(3/2^-)$. For the coupled $\Xi_c^{\prime}\bar{D}/\Xi_c\bar{D}^*/\Xi_c^{\prime}\bar{D}^*/\Xi_c^*\bar{D}^*$ state with $0(1/2^-)$, when the cutoff is taken around 0.8 GeV, the binding energy is around $-10$ MeV, the RMS radius is around 1.0 fm, and the dominant channel is the $S-$wave $\Xi_c^{\prime}\bar{D}$ with its probability around 70\%. If we adopt the former remarks, these bound state solutions are consistent with the reasonable loosely bound state properties. Therefore, we can conclude that this state can be the possible strange hidden-charm molecular pentaquark. Since the probabilities for the remaining channels are almost 30 percent, the coupled channel effects play an important role to generate this coupled bound state. For the isovector coupled $\Xi_c^{\prime}\bar{D}/\Xi_c\bar{D}^*/\Xi_c^{\prime}\bar{D}^*/\Xi_c^*\bar{D}^*$ state with $1/2^-$, the binding energy appears at cutoff $\Lambda$ around 1.10 GeV, the RMS radius is over or around 1.0 fm, and the $S-$wave $\Xi_c^{\prime}\bar{D}$ component is dominant, the corresponding probability is around 90 \%. The reasonable cutoff, the reasonable RMS radius and the small binding energy indicate that the oupled $\Xi_c^{\prime}\bar{D}/\Xi_c\bar{D}^*/\Xi_c^{\prime}\bar{D}^*/\Xi_c^*\bar{D}^*$ state with $1(1/2^-)$ can be the other possible strange hidden-charm molecular candidate.

For the coupled $\Xi_c^{*}\bar{D}/\Xi_c^{\prime}\bar{D}^*/\Xi_c^*\bar{D}^*$ states with $0,1(3/2^-)$, and the coupled $\Xi_c^{\prime}\bar{D}^*/\Xi_c^*\bar{D}^*$ states with $0,1(3/2^-)$, their bound states solutions are similar to the corresponding results for the coupled $\Xi_c^{\prime}\bar{D}/\Xi_c\bar{D}^*/\Xi_c^{\prime}\bar{D}^*/\Xi_c^*\bar{D}^*$ states with $0,1(1/2^-)$. In the range of $1.00<\Lambda<2.00$ GeV, we can obtain the loosely binding energies and the reasonable RMS radii. And the dominant channels are the systems with lowest mass threshold in these discussed coupled systems. Meanwhile, we also find the interactions from the isoscalar states are stronger than those in the isovector states as the corresponding cutoff values are a little smaller than those in the isovector bound states with the same binding energy. If we still adopt the criterions of the loosely bound molecular states, the coupled $\Xi_c^{*}\bar{D}/\Xi_c^{\prime}\bar{D}^*/\Xi_c^*\bar{D}^*$ states with $0,1(3/2^-)$, and the coupled $\Xi_c^{\prime}\bar{D}^*/\Xi_c^*\bar{D}^*$ states with $0,1(3/2^-)$ can be the possible strange hidden-charm molecular candidates.

As shown in Figure \ref{loose}, we also obtain the loosely bound state solutions for the coupled $\Xi_c^{\prime}\bar{D}^*/\Xi_c^*\bar{D}^*$ states with $0(1/2^-)$ when the cutoff $\Lambda$ is taken around 1.00 GeV. The dominant channel is the $S-$wave $\Xi_c^{\prime}\bar{D}^*$ component. Since its probability is over 90 \% in the cutoff range, the coupled channel effects play a minor role in forming this bound state. Therefore, this bound state can be recommended as the possible strange hidden-charm molecular pentaquark.

\begin{figure}[!htbp]
\begin{flushleft}
\includegraphics[width=1.6in]{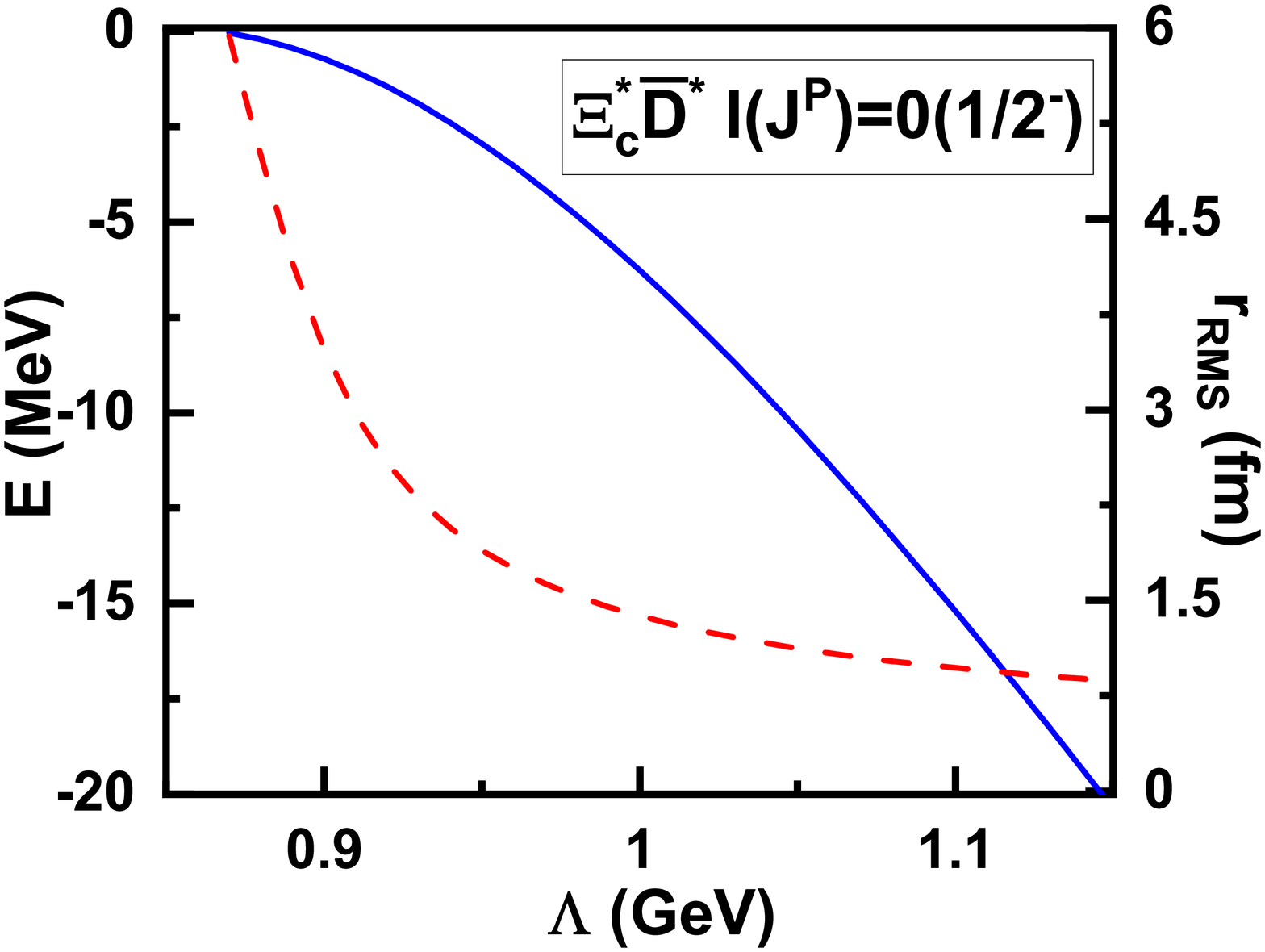}
\includegraphics[width=1.6in]{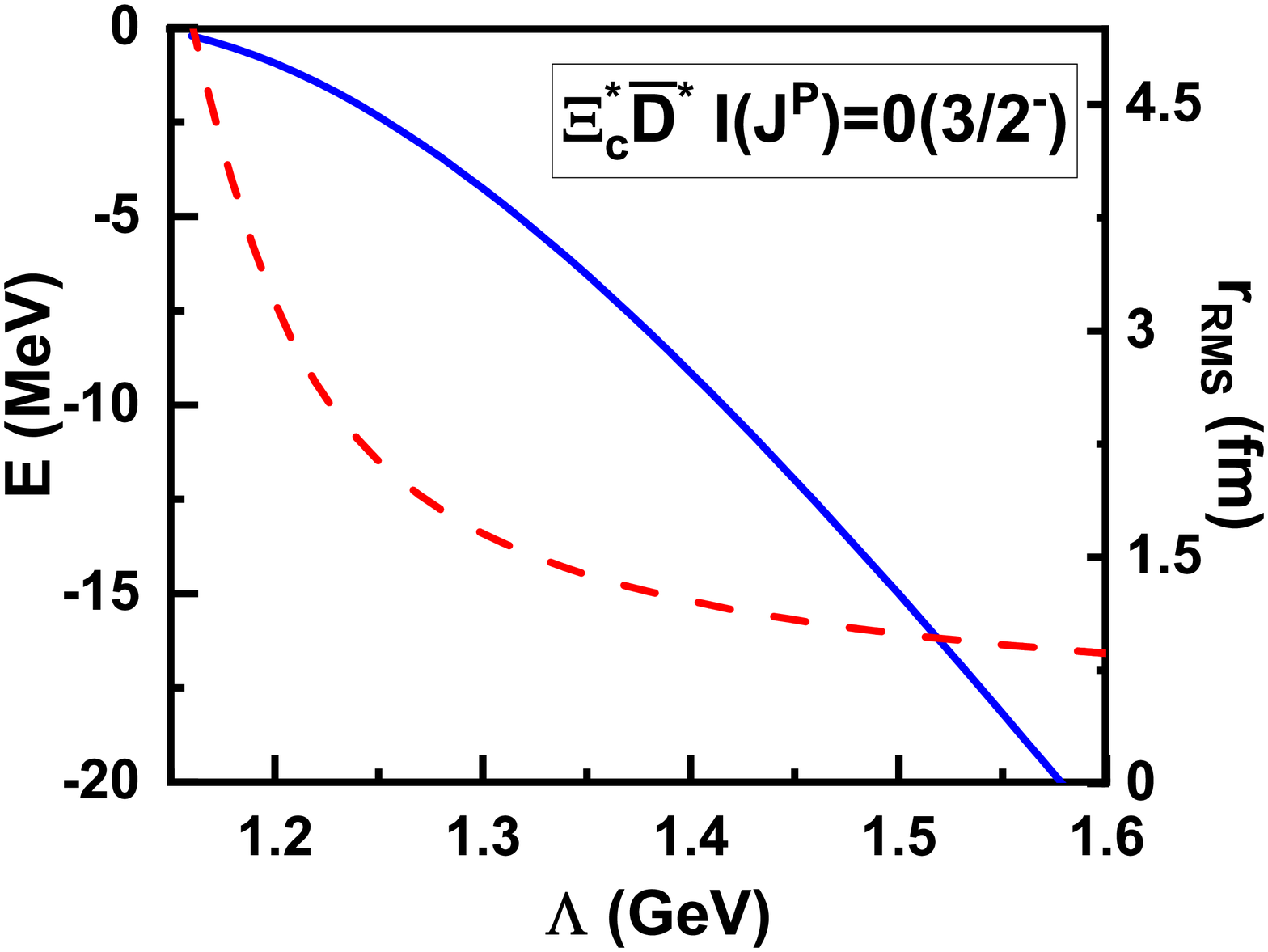}\\
\includegraphics[width=1.6in]{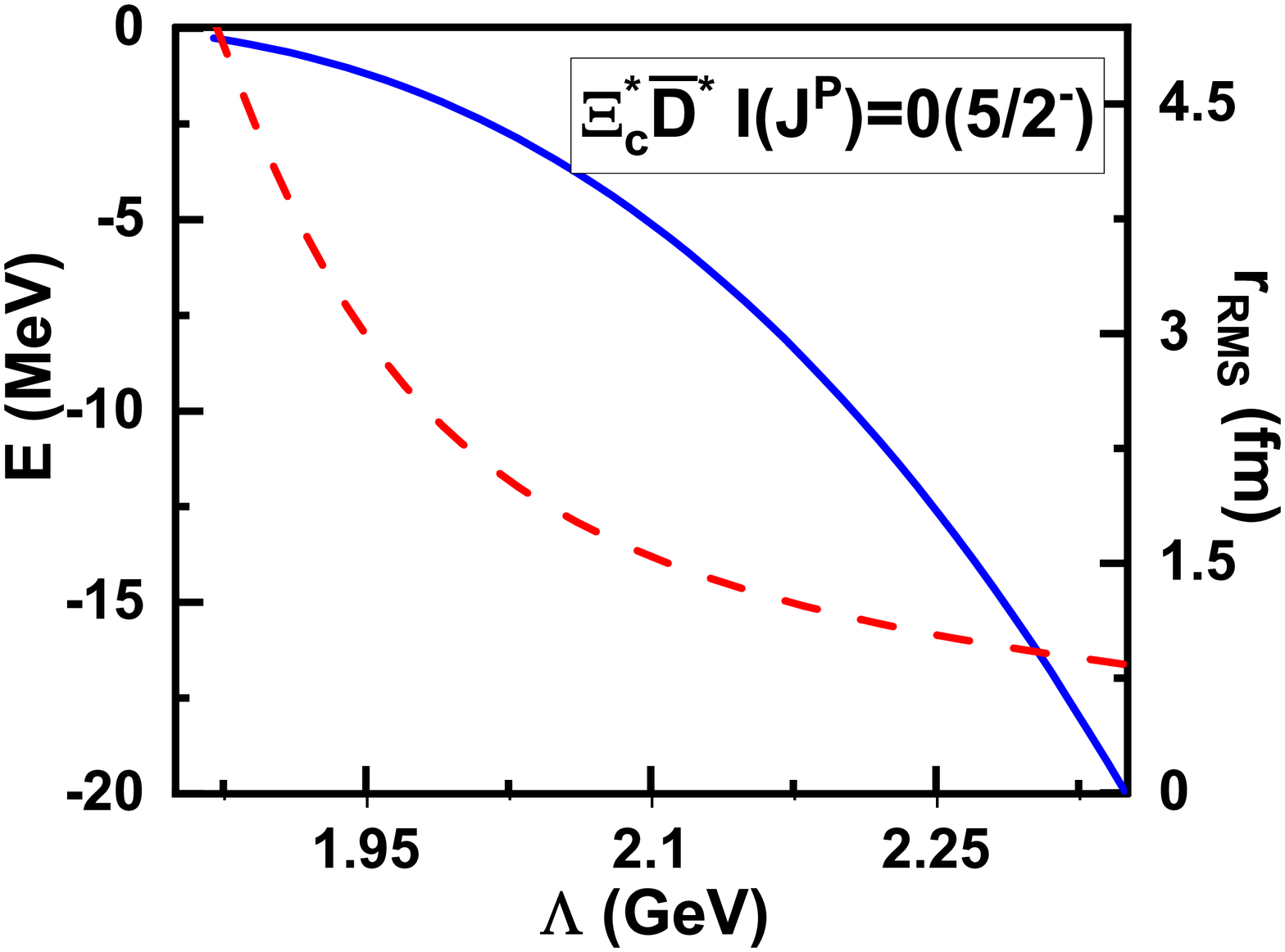}
\includegraphics[width=1.6in]{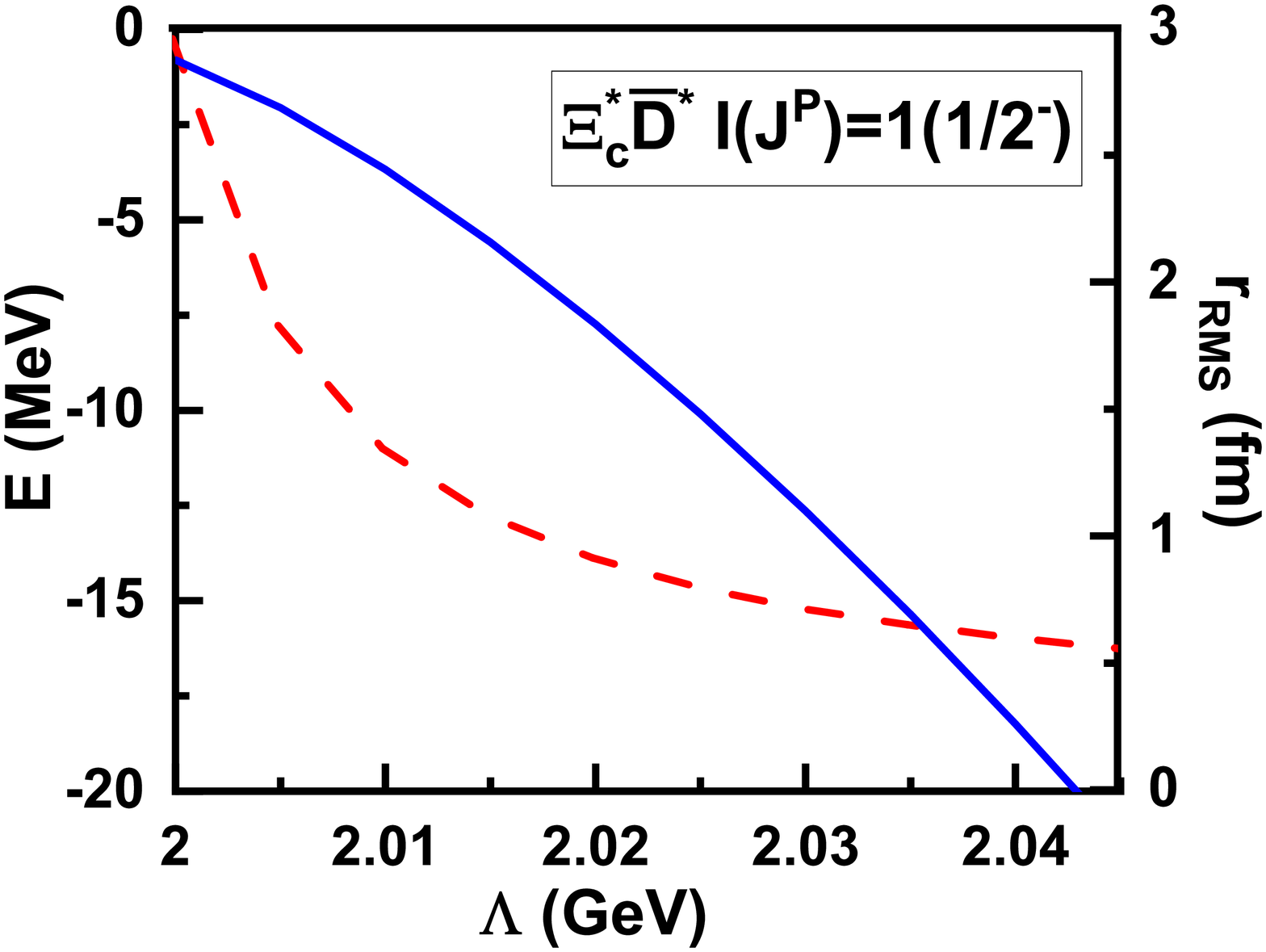}\\
\includegraphics[width=1.6in]{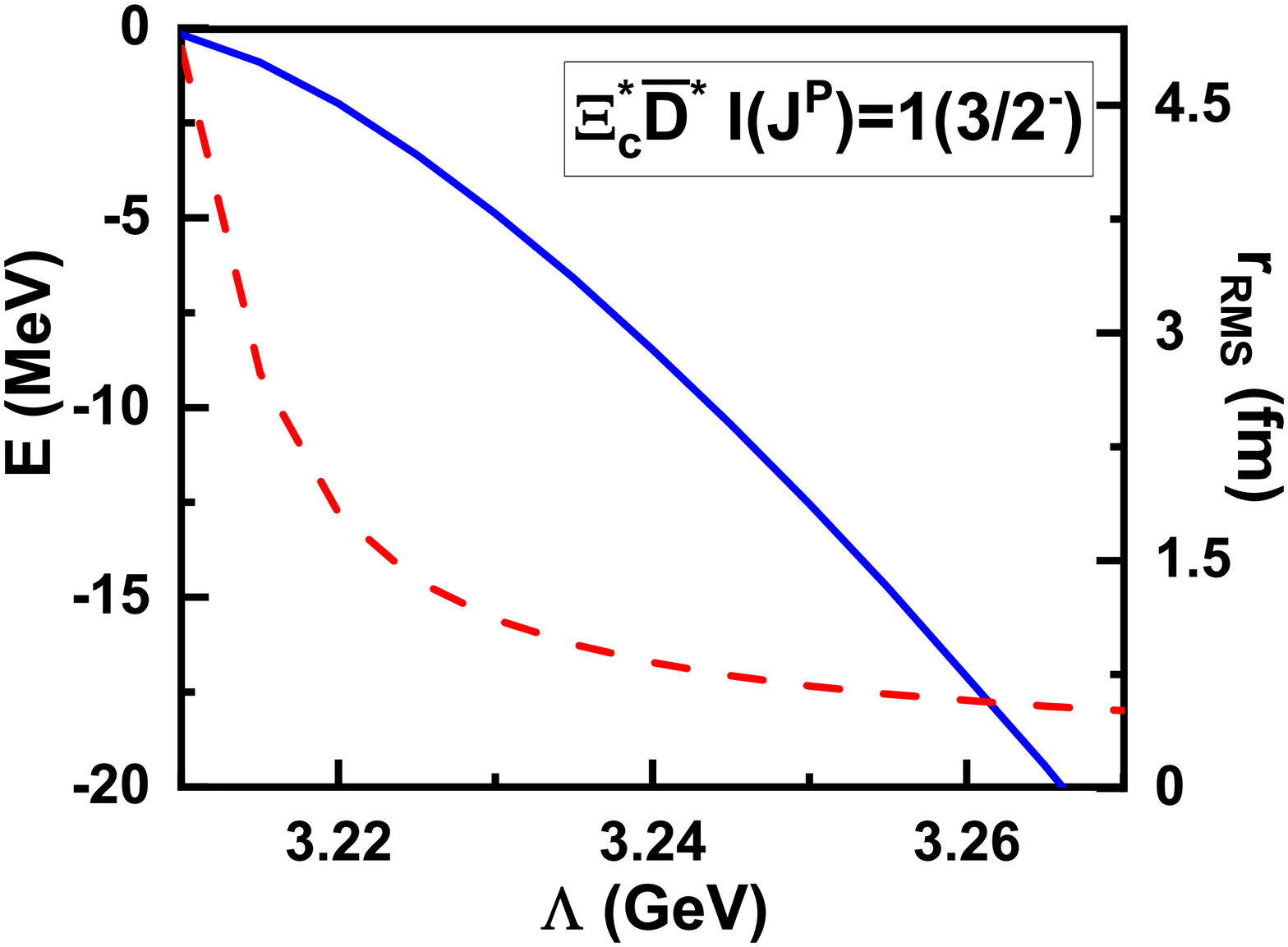}
\end{flushleft}
\caption{The $\Lambda$ dependence of the binding energy $E$ and RMS radius for the single $\Xi_c^*\bar{D}^*$ systems with $I(J^P)=0(1/2^-, 3/2^-, 5/2^-)$ and $1(1/2^-, 3/2^-)$. The blue solid lines and the red dashed lines correspond to the binding energies curves and root-mean-square radii curves, respectively.}\label{xics}
\end{figure}

Meanwhile, we also perform a single $\Xi_c^*\bar{D}^*$ channel analysis to search for possible molecular states. When we vary the cutoff value in the range of 0.80$\sim$5.00 GeV, we find that
\begin{enumerate}
  \item There don't exist bound state solutions for the isovector $\Xi_c^*\bar{D}^*$ system with $5/2^-$.
  \item For the isovector $\Xi_c^*\bar{D}^*$ system with $3/2^-$, it cannot be a good strange hidden-charm molecular candidate as its cutoff $\Lambda$ is larger than 3.00 GeV.
  \item For the $\Xi_c^*\bar{D}^*$ with $1(1/2^-)$ and $0(5/2^-)$, they may the possible strange hidden-charm molecular candidates as their binding energies appear at the cutoff $\Lambda$ around 2.00 GeV.
  \item If we still take the cutoff value $\Lambda$ around 1.00 GeV as a reasonable parameter, the isoscalar $\Xi_c^*\bar{D}^*$ molecules with $J^P=(1/2^-, 3/2^-)$ can be prime strange hidden-charm molecular candidates.
  \item As shown in Figure 2, the OBE interactions become stronger with the increasing of the cutoff value for one bound state.  When we adopt this rough property in the isoscalar single $\Xi_c^*\bar{D}^*$ bound states, one can conclude that the interactions with the higher spin are a little weaker attractive according to the cutoff relation $\Lambda(1/2^-)<\Lambda(3/2^-)<\Lambda(5/2^-)$ with the same binding energy. This is also consist with predictions in Ref. \cite{Wang:2019nvm}.
\end{enumerate}


\renewcommand\tabcolsep{0.3cm}
\renewcommand{\arraystretch}{1.5}
\begin{table*}[!htbp]
\caption{The bound state solutions of the coupled $\Xi_c^{(\prime,*)}\bar{D}^{(*)}$ systems with $I(J^P)=0(1/2^-)$ and $1(1/2^-)$. The cutoff $\Lambda$, the root-mean-square $r_{RMS}$, and the mass of the bound state $M$ are in the units of GeV, fm, and MeV, respectively. $p_i(\%)$ denotes the probability of the $i-$th channel for the investigated system. The largest probability of the quantum number configuration for a bound state are remarked by bold typeface. And $\tilde E$, in the unit of MeV, is the mass gap between the bound state mass and the threshold of the dominant channel.}\label{xicd}
\begin{tabular}{lc|rrr|rrr|rrr}
\toprule[2pt]
\multicolumn{2}{c|}{$I(J^P)$}
&\multicolumn{3}{c|}{${0({1}/{2}^-)}$}    &\multicolumn{3}{c|}{${1({1}/{2}^-)}$}
 &\multicolumn{3}{c}{${1({1}/{2}^-)}$}
\\\hline
$\Lambda$
 &    &0.99      &1.00       &1.01
     &1.27      &1.28       &1.29
     &2.03      &2.04       &2.05\\
$r_{RMS}$
 &    &1.29      &0.60      &0.49
     &0.40      &0.36      &0.34
     &0.89      &0.40      &0.34\\
 $M$
&    &4334.64    &4327.69   &4319.82
    &4327.53   &4315.33      &4302.35
    &4583.26   &4574.51      &4565.02\\

$\tilde E$    & &$-$109.37       &$-$116.32     &$-$124.19
&$-$116.48       &$-$128.68     &$-$141.66
&$-$71.2       &$-$79.95     &$-$89.44
\\
$p_i(\%)$
&$\Xi_c\bar{D}$     &30.36       &16.51     &12.37
    &2.12        &1.48     &1.19
    &$\ldots$        &$\ldots$     &$\ldots$\\

&$\Xi_c^{\prime}\bar{D}$     &\bf 32.56       &\bf 38.99     &\bf 40.73
     &\bf 62.36       &\bf 62.33     & \bf 62.10
    &$\ldots$        &$\ldots$     &$\ldots$\\

&$\Xi_c\bar{D}^*$     &21.75       &26.38     &27.89
&35.30       &35.97     &36.50
    &$\ldots$        &$\ldots$     &$\ldots$
\\

&$\Xi_c^{\prime}\bar{D}^*$     &13.07       &15.58     &16.40
&0.17       &0.18     &0.19
&16.33       &8.57     &6.91
\\

&$\Xi_c^{*}\bar{D}^*$     &2.26       &2.54     &2.61
&0.04       &0.03     &0.02
&\bf 83.67       &\bf 91.43     &\bf 93.09
\\
\bottomrule[2pt]
\end{tabular}
\end{table*}

In our calculations, we also obtain the other kind of bound state solutions, which is very different with those for a reasonable loosely bound hadronic molecule. In this case, their RMS radii are around 0.5 fm or more less, therefore they cannot be reasonable hadronic molecular states but the tightly bound states. According to the relation $R\sim 1/\sqrt{2\tilde{\mu} \tilde{E}}$, the system with a higher mass is always the dominant channel for the tightly bound state.

As shown in Table \ref{xicd}, we obtain three tightly bound state solutions, the coupled $\Xi_c\bar{D}/\Xi_c^{\prime}\bar{D}/\Xi_c\bar{D}^*/\Xi_c^{\prime}\bar{D}^*/\Xi_c^*\bar{D}^*$ states with $I(J^P)=0(1/2^-)$ and $1(1/2^-)$ and the coupled $\Xi_c^{\prime}\bar{D}^*/\Xi_c^*\bar{D}^*$ state with $1(1/2^-)$, where the other binding energies $\tilde E=E+(M_{\text{lowest}}-M_{\text{dominant}})$ are around 100 MeV, the corresponding dominant channels are $\Xi_c^{\prime}\bar{D}$ and $\Xi_c^*\bar{D}^*$, respectively. Compared to the coupled $\Xi_c\bar{D}/\Xi_c^{\prime}\bar{D}/\Xi_c\bar{D}^*/\Xi_c^{\prime}\bar{D}^*/\Xi_c^*\bar{D}^*$ states with $I(J^P)=0(1/2^-)$ and $1(1/2^-)$, the cutoff value in the coupled $\Xi_c^{\prime}\bar{D}^*/\Xi_c^*\bar{D}^*$ state with $1(1/2^-)$ is a little away from the empirical value $\Lambda$ around 1.00 GeV.

For the $\Xi_c\bar{D}/\Xi_c^{\prime}\bar{D}/\Xi_c\bar{D}^*/\Xi_c^{\prime}\bar{D}^*/\Xi_c^*\bar{D}^*$ interactions, if we recall that such interactions are obtained by adding the lowest channel from the $\Xi_c^{\prime}\bar{D}/\Xi_c\bar{D}^*/\Xi_c^{\prime}\bar{D}^*/\Xi_c^*\bar{D}^*$ interactions, it is not strange to find the analogous results in the $\Xi_c^{\prime}\bar{D}/\Xi_c\bar{D}^*/\Xi_c^{\prime}\bar{D}^*/\Xi_c^*\bar{D}^*$ interactions. Thus, the $\Xi_c\bar{D}/\Xi_c^{\prime}\bar{D}/\Xi_c\bar{D}^*/\Xi_c^{\prime}\bar{D}^*/\Xi_c^*\bar{D}^*$ tightly bound states with $0,1(1/2^-)$ here are not the independent states, but correspond to the $\Xi_c^{\prime}\bar{D}$ loosely bound molecular pentaquarks with $0,1(1/2^-)$.

\section{conclusion and discussion}\label{sec4}

As a hot issue in the hadron physics, whether the new hadron states are the hadronic molecules is still open to discuss. In 2019, the discovery of the three $P_c$ states could provide a strong evidence of the existence of the hidden-charm molecular pentaquarks \cite{Aaij:2019vzc}. In Ref. \cite{Chen:2019asm}, the $P_c(4312)$, $P_c(4440)$, and $P_c(4457)$ states can be assigned as the $\Sigma_c\bar{D}$ molecular state with $1/2(1/2^-)$, the $\Sigma_c\bar{D}^*$ states with $1/2(1/2^-)$ and $1/2(3/2^-)$, respectively. This is not the end of the story. Very recently, the LHCb Collaboration reported a new evidence of the strange hidden-charm pentaquark $P_{cs}(4459)$ in the $\Xi_b^-\to J/\psi\Lambda K^-$ process. Theorists also proposed that the $P_{cs}(4459)$ can be the strange hidden-charm molecular state composed of the $S-$wave $\Xi_c\bar{D}^*$ state \cite{Chen:2020kco,Chen:2021tip,Chen:2020uif,Peng:2020hql,Wang:2020eep}.

\begin{figure}[!htbp]
\center
\includegraphics[width=3.4in]{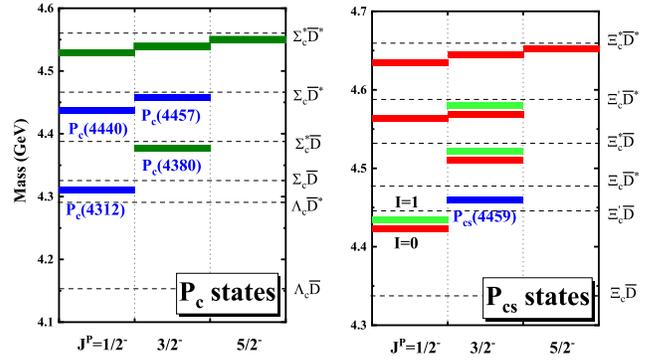}
\caption{A summary of the mass spectrum of the $P_c$ states in the meson-baryon molecular scenario \cite{Chen:2019asm} and the predicted possible hidden-charm molecular pentaquarks candidates with $|S|=1$. Here, we roughly estimate the mass positions of the predicted hidden-charm molecular pentaquarks according to the values of the cutoff $\Lambda$. The red and green lines label the predicted molecular candidates with $I=0$ and $I=1$, respectively.}\label{prediction}
\end{figure}

In this work, we preform the coupled channel analysis on the interactions between a charm-strange baryon $\Xi_c^{(',*)}$ and an anti-charmed meson $\bar{D}^{(*)}$ in the framework of the OBE model. By using the $SU(3)$ flavor symmetry and the heavy quark symmetry, we can obtain the OBE effective potentials relations between the $\Xi_c^{(\prime,*)}\bar{D}^{(*)}$ interactions and the $\Sigma_c^{(*)}\bar{D}^{(*)}$ interactions, where we have already deduced the concrete OBE effective potentials in Ref. \cite{Chen:2019asm}. For the $\Xi_c\bar{D}^{(*)}$ systems, their $SU(3)$ flavor partners are the $\Lambda_c\bar{D}^{(*)}$ systems, which is not prepared in our former work. In this work, we derive the corresponding OBE effective potentials in the general procedures.

Our results can predict several possible hidden-charm molecular pentaquarks with strangeness $|S|=1$. As shown in the Figure \ref{prediction}, there can exist seven possible isoscalar hidden-charm molecular pentaquarks with $|S|=1$, they are mainly composed of the $\Xi_c^{\prime}\bar{D}$ state with $I(J^P)=0(1/2^-)$, the $\Xi_c\bar{D}^*$ state with $0(3/2^-)$, the $\Xi_c^*\bar{D}$ state with $0(3/2^-)$, the $\Xi_c^{\prime}\bar{D}^*$ states with $0(1/2^-, 3/2^-)$, and the $\Xi_c^*\bar{D}^*$ states with $0(1/2^-, 3/2^-)$, respectively. The $\Xi_c^*\bar{D}^*$ state with $0(5/2^-)$ may be also the possible strange hidden-charm molecular candidate. Meanwhile, we find there may exist three isovector hidden-charm molecular pentaquarks with $|S|=1$, like the $\Xi_c^{\prime}\bar{D}$ state with $1(1/2^-)$, the $\Xi_c^*\bar{D}$ state with $1(3/2^-)$, and the $\Xi_c^{\prime}\bar{D}^*$ state with $1(3/2^-)$. Our results also indicate that the coupled channel effect play a very important role in forming these hidden-charm molecular candidates with strangeness $|S|=1$, especially for the $\Xi_c^{\prime}\bar{D}/\Xi_c\bar{D}^*/\Xi_c^{\prime}\bar{D}^*/\Xi_c^*\bar{D}^*$ coupled states with $I(J^P)=0,1(1/2^-)$, the $\Xi_c^{*}\bar{D}/\Xi_c^{\prime}\bar{D}^*/\Xi_c^*\bar{D}^*$ coupled state with $0(3/2^-)$, and the $\Xi_c^{\prime}\bar{D}^*/\Xi_c^*\bar{D}^*$ coupled states with $0,1(3/2^-)$. We are looking forward that the future experiments can search for possible hidden-charm pentaquarks with strangeness around these predicted mass thresholds.

Compared to the $P_c$ states assigned to the meson-baryon molecules, as shown in Figure \ref{prediction}, there are four more possible hidden-charm molecular pentaquarks with strangeness $|S|=1$, which include one isoscalar $\Xi_c\bar{D}^*$ state with $3/2^-$ and three isovector bound states. In our previous work \cite{Chen:2020kco,Chen:2020uif}, the $P_{cs}(4459)$ can be explained as the isoscalar $\Xi_c\bar{D}^*$ state with $3/2^-$, and the $\Xi_c^*\bar{D}$ channel is also very important. As we seen, the mass difference between the $\Lambda_c\bar{D}^*$ and $\Sigma_c\bar{D}^*$ systems is much larger than that between the $\Xi_c\bar{D}^*$ and $\Xi_c^*\bar{D}$ systems, which can weaken the contribution from the coupled channel effect, this may explain the reason why there cannot exist the possible hidden-charm molecular pentaquarks composed of the $\Lambda_c\bar{D}^*$ state with $1/2(3/2^-)$.

As shown in Table \ref{potential}, the OBE interactions for the isovector $\Xi_c^{(',*)}\bar{D}^{(*)}$ systems are weaker repulsive or stronger attractive than those from the $\Sigma_c^{(*)}\bar{D}^{(*)}$ systems with $I=3/2$. This is the main reason that we may predict three isovector strange hidden-charm molecular pentaquarks instead of the hidden-charm molecular pentaquarks with $I=3/2$ \cite{Chen:2019asm}.

\section*{ACKNOWLEDGMENTS}

Rui Chen is very grateful to Shi-Lin Zhu for helpful discussions and constructive suggestions. This work is supported by the China National Funds for Distinguished Young Scientists under Grant No. 11825503 and by the National Program for Support of Top-notch Young Professionals. R. C. is supported by the National Postdoctoral Program for Innovative Talent.

\end{document}